# Formation of Venus, Earth and Mars: Constrained by isotopes

Helmut Lammer[1], Ramon Brasser[2], Anders Johansen[3], Manuel Scherf[1], Martin Leitzinger[4]

[1]Space Research Institute, Austrian Academy of Sciences, Schmiedlstr. 6, 8042 Graz, Austria

[2]Earth-Life Science Institute, Tokyo Institute of Technology, 2-12-1 Ookayama, Meguro-ku, Tokyo 152-8551 Japan

[3]Lund Observatory, Department of Astronomy and Theoretical Physics, Lund University, Box 43, 22100 Lund Sweden

[4]Institute of Physics/IGAM, University of Graz, Universitätsplatz 5/II, 8010 Graz, Austria

**Abstract:** Here we discuss the current state of knowledge of terrestrial planet formation from the aspects of different planet formation models and isotopic data from $^{182}$Hf-$^{182}$W, U-Pb, lithophile-siderophile elements, $^{48}$Ca/$^{44}$Ca isotope samples from planetary building blocks, recent reproduction attempts from $^{36}$Ar/$^{38}$Ar, $^{20}$Ne/$^{22}$Ne, $^{36}$Ar/$^{22}$Ne isotope ratios in Venus' and Earth's atmospheres, the expected solar $^3$He abundance in Earth's deep mantle and Earth's D/H sea water ratios that shed light on the accretion time of the early protoplanets. Accretion scenarios that can explain the different isotope ratios, including a Moon-forming event ca. 50 Myr after the formation of the Solar System, support the theory that the bulk of Earth's mass ($\geq$ 80 %) most likely accreted within 10 - 30 Myr. From a combined analysis of the before mentioned isotopes, one finds that proto-Earth accreted a mass of 0.5 – 0.6 $M_{\text{Earth}}$ within the first ≈ 4-5 Myr, the approximate lifetime of the protoplanetary disk. For Venus, the available atmospheric noble gas data are too uncertain for constraining the planet's accretion scenario accurately. However, from the available imprecise Ar and Ne isotope measurements, one finds that proto-Venus could have grown to a mass of 0.85 – 1.0 $M_{\text{Venus}}$ before the disk dissipated. Classical terrestrial planet formation models have struggled to grow large planetary embryos, or even cores of giant planets, quickly from the tiniest materials within the typical lifetime of protoplanetary disks. Pebble accretion could solve this long-standing time scale controversy. Pebble accretion and streaming instabilities produce large planetesimals that grow into Mars-sized and larger planetary embryos during this early accretion phase. The later stage of accretion can be explained well with the Grand-Tack model as well as the annulus and depleted disk models. The relative roles of pebble accretion and planetesimal accretion/giant impacts are poorly understood and should be investigated with N-body simulations that include pebbles and multiple protoplanets. To summarise, different isotopic dating methods and the latest terrestrial planet formation models indicate that the accretion process from dust settling, planetesimal formation, and growth to large planetary embryos and protoplanets is a fast process that occurred to a great extent in the Solar System within the lifetime of the protoplanetary disk.

# 1. Introduction

Terrestrial planets accrete mass by numerous collisions between small objects that accumulate into larger ones, thereby forming planetesimals, and Moon- and Mars-mass planetary embryos that eventually grow into planets. An illustration of a schematic time scale of the accretion of terrestrial planets is shown in Fig. 1. One can see four main accretion stages: i) dust settling; ii) planetesimal formation; iii) formation of Moon- to Mars-mass planetary embryos; and iv) the accretion of protoplanets up to terrestrial planetary masses by giant impacts (e.g., Morbidelli et al., 2012; Brasser, 2013; Johansen et al., 2015; Birnstiel et al., 2016). The bulk composition of the silicate part of terrestrial planets has long been linked to a fraction of carbonaceous chondritic meteorites (CI, CV, CM, etc.), which was mixed with enstatite chondrites (EC) and ordinary chondrites (OC) according to mainly oxygen isotope evidence (Drake and Righter, 2002; Dauphas, 2017; Birmingham et al., 2020, this issue), and/or ureilite-like material based on recent analysed $^{48}Ca/^{44}Ca$ data in samples that originated from ureilite and angrite meteorites, as well as from Vesta, Mars and Earth (Schiller et al., 2018).

As illustrated in Fig. 1, depending on their speed of growth before or after disk dissipation and their orbital locations of origin/destination, protoplanets will evolve into objects with different volatile and elemental abundances, which results in different elemental ratios compared to their initial ones due to subsequent atmospheric escape and/or impact erosion of atmospheres and crustal/mantle material (e.g., Lammer et al., 2020a; this issue; O'Neill et al., 2020; this issue).

Safronov (1969) proposed the first physical model for the formation of the terrestrial planets. He suggested that the earliest stage of accumulation of dust into planetesimals was caused by gravitational instabilities in a thin dust layer in the midplane of the protoplanetary disk. Due to their gravitational interaction, the ensuing planetesimals are stirred and can encounter each other. Safronov (1969) showed that the relative velocities between planetesimals are typically of the order of their surface escape velocity, so that each planetesimal's gravitational cross section is limited by the geometrical one, restricting the accretion rate. Wetherill (1980) built upon these ideas and suggested that the terrestrial planets coagulated from planetesimals, and that the formation of these planets was linked with the evolution of the asteroid belt. Wetherill and Stewart (1989) subsequently showed that in a disk of planetesimals, some would undergo runaway growth and form a sequence of planetary embryos. These embryos would then further collide to form the terrestrial planets.

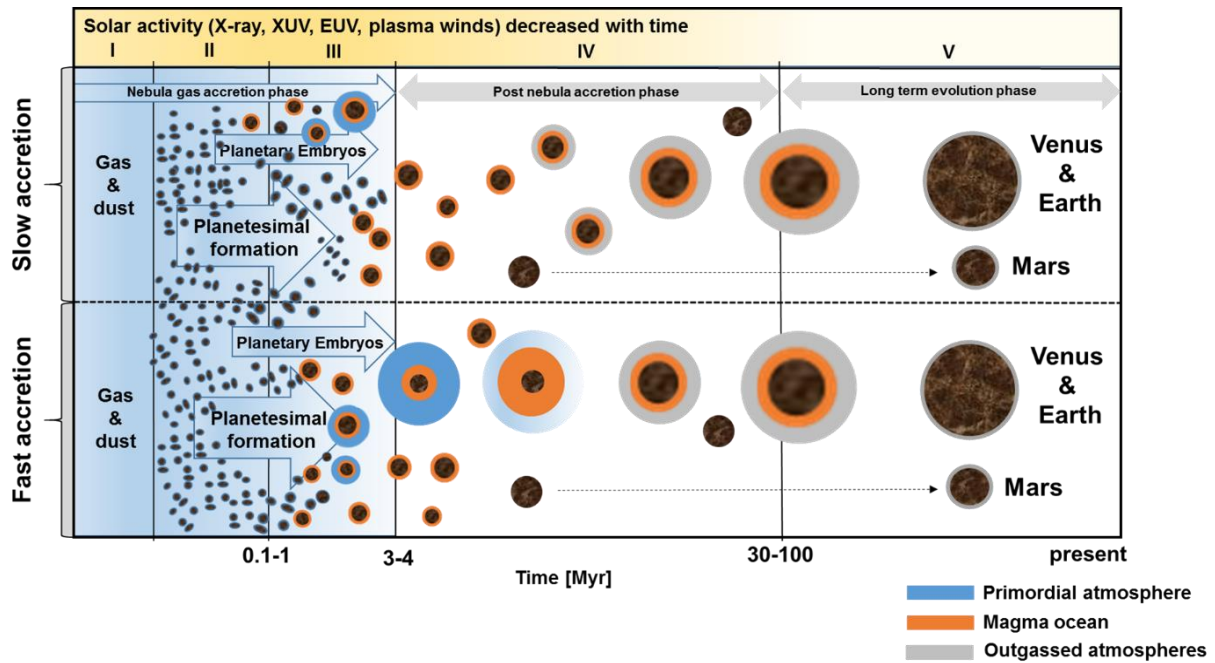

**Fig. 1**: Illustration of the five main formation stages of the terrestrial planet formation: dust settling (I), planetesimal formation (II), formation of planetary embryos (III), giant impacts including Earth's Moon-forming event (IV). The long-term evolution phase where secondary atmospheres origin (V). The scenarios above the horizontal dashed line illustrate so-called slow accretion scenarios where terrestrial planets accrete most of their mass after the disk dissipated. The scenario below the dashed-horizontal line corresponds to a two-stage accretion scenario according to Jacobsen and Harper (1996), Harper and Jacobsen (1996), and Yu and Jacobsen (2011). Here, terrestrial planets accrete faster so that they can capture primordial atmospheres, which subsequently escape while the proto-planets accrete their final masses. The shaded zone represents the presence of the protoplanetary nebula in the Solar System that dissipated after ≈ 3.3 – 4.5 Myr (Bollard et al., 2017; Wang et al., 2017). In both scenarios, the planets outgas steam atmospheres when their magma oceans solidify, and secondary atmospheres build up later-on due to volcanic outgassing and tectonic activities.

The past decades have seen much progress in understanding the formation of the terrestrial planets. Models that fully explain the formation of terrestrial planets in the inner Solar System have to explain how planetesimals grow to ≈ 100 km-scale bodies because of dust coagulation (Johansen et al., 2017). These planetesimals then further coagulate into Mars-mass or even larger planetary embryos, most likely by planetesimal accretion (Kokubo and Ida, 1995; Kokubo and Ida, 1998). Finally, terrestrial planet formation models that form Venus, Earth and Mars must match constraints related to the involved planetary embryos and asteroids, including their orbital and compositional distributions as well as their isotope and elemental ratios (Morbidelli et al., 2012; Brasser et al., 2017).

At present, there exists a great variety of terrestrial planet formation models: the Classical model (e.g., Wetherill and Stewart, 1989; Chambers, 2001; Raymond et al., 2006; 2009), Grand Tack model (e.g., Walsh et al., 2011; Brasser et al., 2017; 2018, Woo et al., 2018), annulus model (Agnor et al. 1999; Hansen, 2009; Walsh and Levison, 2016), depleted disk model (e.g., Izidoro et al., 2014; 2015; Raymond and Izidoro, 2017a) and early instability model (Clement et al., 2019). In contrast, giant planet formation in the solar system has not been studied as extensively (Helled et al., 2013). As such, most of the terrestrial planet formation models in the literature presuppose that terrestrial and giant planet formation can be treated separately because they frequently assume that the giant planets have already formed and reside on their current orbits. Only the Grand Tack model includes the migration of the giant planets. The models listed above tend to study the later stages of terrestrial planet formation, usually focused on the giant impact stage (Chambers, 2001) that follows the oligarchic growth phase (Kokubo and Ida, 1998). However, we would argue that an additional model of terrestrial planet formation can be constructed based on pebble accretion (e.g., Ormel and Klahr, 2010; Lambrechts and Johansen, 2012; Lambrechts et al., 2014; Drazkowska et al., 2016). Even though these studies focus primarily on the formation of planets from a stream of pebbles spiralling through the disc and accreting onto a planetary core, Levison et al., (2015b) constructed a model of the formation of the terrestrial planets from pebble accretion starting from a planetesimal disk and then ran this all the way through the oligarchic and giant impact phases. Their model suggests that the oligarchic and giant impact phases are a natural outcome irrespective of how the planetesimals are formed and what they subsequently accrete. The only potential difference is the time scale of accretion.

Here we discuss most of the terrestrial planet formation models in varying degrees of depth (Sect. 2). In Sect. 3, we summarise the latest knowledge on terrestrial planet accretion, including time scales that are obtained from data based on the isotopic chronometers $^{182}$Hf-$^{182}$W and U-Pb, lithophile-siderophile elemental isotope data of accreted material, the latest data analysis from $^{48}$Ca/$^{44}$Ca isotope samples from planetary building blocks, recent reproduction attempts of Venus' and Earth's atmospheric $^{36}$Ar/$^{38}$Ar, $^{20}$Ne/$^{22}$Ne, $^{36}$Ar/$^{22}$Ne isotope ratios, the evidence of solar $^{3}$He abundance in Earth's deep mantle, and the D/H ratio in Earth's sea water. These findings are then compared with the various terrestrial planet formation models. Finally, Sect. 4 concludes the review.

## 2. Terrestrial planet formation models

### 2.1 The Classical model for terrestrial planet formation

Wetherill and Stewart (1989) demonstrated that in a protoplanetary disk consisting of equal-mass planetesimals, some would undergo rapid collisions with other planetesimals leading to a phase of runaway growth wherein the mass of the planet increases as $dM/dt \propto M^{\beta}$, where $\beta$ is a free parameter. When $\beta > 1$ runaway growth occurs. In contrast, if $\beta < 1$ then the accretion is more egalitarian and no specific body will begin to dominate over the others.

During the earliest stages of planet formation $\beta=4/3$ due to effective gravitational focusing (Ida & Makino, 1993; Kokubo and Ida, 1996). Here the planetesimals gravitationally stir each other (Ida and Makino, 1993), resulting in their eccentricities and inclinations slowly increasing. Once the eccentricities reach orbit-crossing values, the planetesimals will eventually collide. Under the assumption of perfect mergers, the formation enters a runaway growth phase in which the size-frequency distribution of the bodies in the disk becomes approximately bimodal (Kokubo and Ida, 1996). During this phase the solids in the disk consist of approximately lunar- to Mars-sized protoplanets (dubbed planetary embryos) spaced by about 10 mutual Hill radii, which are embedded in a sea of planetesimals (Kokubo and Ida, 1998). The Hill radius of a planet is given by $R_H = a(m_{pl}/3M_{star})^{1/3}$, where $a$ is the semi-major axis, $m_{pl}$ and $M_{star}$ are the mass of the planet and the star; the mutual Hill radius uses the sum of the masses of two neighbouring planets and the mean of their semi-major axes. During the bimodal phase the planetary embryos gravitationally perturb the planetesimals and each other (Kokubo and Ida, 1998). If no gas remains from the protoplanetary disk the system would be unstable, but if any residual gas remains then the eccentricities and inclinations of all bodies are kept low due to damping forces from the gas, and the system is dynamically quasi-stable (Kominami and Ida., 2002). The planetary embryos all grow at roughly the same rate as they vie for space and for their ability to accrete planetesimals. This phase is called the oligarchic growth stage, and approximately half of the solid mass within the disk is in the embryos, with the other half being in planetesimals (Kokubo and Ida, 1998).

The gas density from the solar nebula decreases with time (e.g. Hartmann et al., 1998) so that the damping forces from the gas disk on the embryos also weaken with time. There comes a point in time when gravitational stirring overcomes the tidal forces from the gas disk (Kominami and Ida, 2002; 2004; Kominami et al. 2005). By then, the mutual gravitational perturbations from the embryos cause their eccentricities to rise to orbit-crossing values so that the whole chain eventually becomes dynamically unstable and the system subsequently evolves into a giant impact phase. Here planetary mergers through large impacts are commonplace. Over tens of millions of years these mutual protoplanetary collisions eventually lead to the observed terrestrial planet inventory (e.g. Chambers and Wetherill, 1998; Chambers, 2001; Raymond et al., 2006, 2009).

The so-called Classical model is based on the evolution described above. Due to limitations in computing power, however, the runaway growth was extremely difficult to model so that simulations of this model began with the Sun being surrounded by a disk of planetesimals and protoplanets that are gravitationally interacting (e.g. Raymond et al., 2006). Typical simulations of the Classical model assume that the gas from the protoplanetary disk has already dissipated (Chambers, 2001), and that Jupiter and Saturn have already formed and reside on their current orbits. More recent iterations of invoke the presence of gas from the protoplanetary disk due to advances in the theory of planet migration and eccentricity damping (e.g. Tanaka et al., 2002; Tanaka and Ward, 2004; Creswell and Nelson, 2008; Paardekooper et al., 2011). In N-body simulations of the model, the protoplanets are usually assumed to have the same size and mass; the same is true for the planetesimals. The surface density of material typically has a steep gradient (typically $\Sigma \propto r^{-3/2}$) to match the mass distribution of the planets (Weidenschilling, 1977).

Early simulations based on the Classical model of terrestrial planet formation estimates the growth time scale of the terrestrial planets to be tens of millions of years. Overall, numerical simulations indicated that the final terrestrial system would be assembled within 100 Myr (Chambers, 2001; Raymond et al., 2006; 2009). In most of these early simulations, however, the resulting planetary systems were found to suffer from excess eccentricity and inclination of the final planets, but the inclusion of a large number of planetesimals – which exert strong dynamical friction on the growing planets – alleviated this concern (O'Brien et al., 2006). Still, a chronic and fundamental shortcoming of the Classical model remained: the output systematically yielded a Mars analogue that was far too massive (e.g. Chambers, 2001; Raymond et al., 2006; 2009).

### 2.2 The Grand Tack model

The problem of Mars' low mass compared to Earth's has been the focus of several recent models of terrestrial planet formation. The most common of these is the Grand Tack model (Walsh et al., 2011), which relies on the early, gas-driven radial migration of Jupiter. The idea behind the model is that after the gas giant formed it opened an annulus in the gas disk and continued migrating towards the Sun (Lin and Papaloizou, 1986) because the torques acting on the planet from the protoplanetary disk are imbalanced. At the same time, Walsh et al. (2011)

argue that beyond Jupiter Saturn also accreted its gaseous envelope, albeit more slowly. Once Saturn reached some critical mass (as suggested by Walsh et al. (2011) to be $50M_{Earth}$, but it could be lower) it rapidly migrated Sunwards too, caught up with Jupiter and became trapped in the 2:3 mean motion resonance (Masset and Snellgrove, 2001). This particular, and fortuitous, configuration of orbital spacing and mass ratio between the gas giants reversed the total torque from the gas disk on both planets (Masset and Snellgrove, 2001), thus reversing their migration; the planets 'tacked', a term from sailing reflecting a change of direction. Pierens et al. (2014) showed that outward migration is possible with Jupiter and Saturn in both the 2:3 and 1:2 resonances (cf. Zhang and Zhou, 2010), but that dynamically stable evolution with little migration may also happen i.e. the planets could stall once they are caught in a resonance. Thus, the inward-outward migration is a possible outcome, but is not guaranteed, and the amount of outward migration depends sensitively on the disk structure and evolution (D'Angelo & Marzari, 2012). To ensure that Mars did not grow beyond its current mass, the location of the 'tack' of Jupiter was set at 1.5 AU (Walsh et al., 2011). Apart from Mars' low mass compared to Earth, the Grand Tack scenario can further account for the semi major axis distribution of the terrestrial planets and the stirred demography of the asteroid belt (DeMeo and Carry, 2014; Brasser et al., 2019).

The dynamical consequences of the Grand Tack model on the terrestrial planet region are two-fold. First, as Jupiter migrated towards the Sun, it strongly scattered all protoplanets and planetesimals in its path due to its high mass. This scattering places more than half of the mass of the disk from the inner Solar System to farther than 5 AU from the Sun. Second, Jupiter pushed some protoplanets and planetesimals towards the Sun. This shepherding of material mostly occurred within the 2:1 mean motion resonance with the planet. When Jupiter reversed its migration at 1.5 AU it caused a pileup of solid material within 1 AU. Thus, almost all protoplanets and planetesimals that originally formed beyond 1 AU were scattered outwards by Jupiter, with a few pushed within 1 AU. This evolution provides a plausible explanation for the low amount of mass in the asteroid belt region.

Figure 2 shows an example of the evolution of a Grand Tack simulation taken from Brasser et al. (2016). The top-left panel depicts the initial conditions. Jupiter and Saturn are at 3.5 and 4.5 AU, respectively. The protoplanetary disk resides between 0.7 AU and 8 AU, interrupted by the gas giants. The disk contains planetary embryos of 0.05 Earth masses and two thousand planetesimals.

Closer than 1.5 AU the disk is assumed to consist of material similar to enstatite chondrite (EC; green) (Brasser et al., 2017; 2018), while between 1.5 AU and Jupiter the material is akin to ordinary chondrite (OC; orange). The blue dots correspond to the carbonaceous chondrites (CC). Symbol sizes are proportional to their masses; for Jupiter and Saturn a fixed size was chosen so as not to fill most of the panels As the gas giants migrate inwards (top-right panel) a significant fraction of OC and CC material is ejected while some OC material is pushed closer to the Sun by Jupiter.

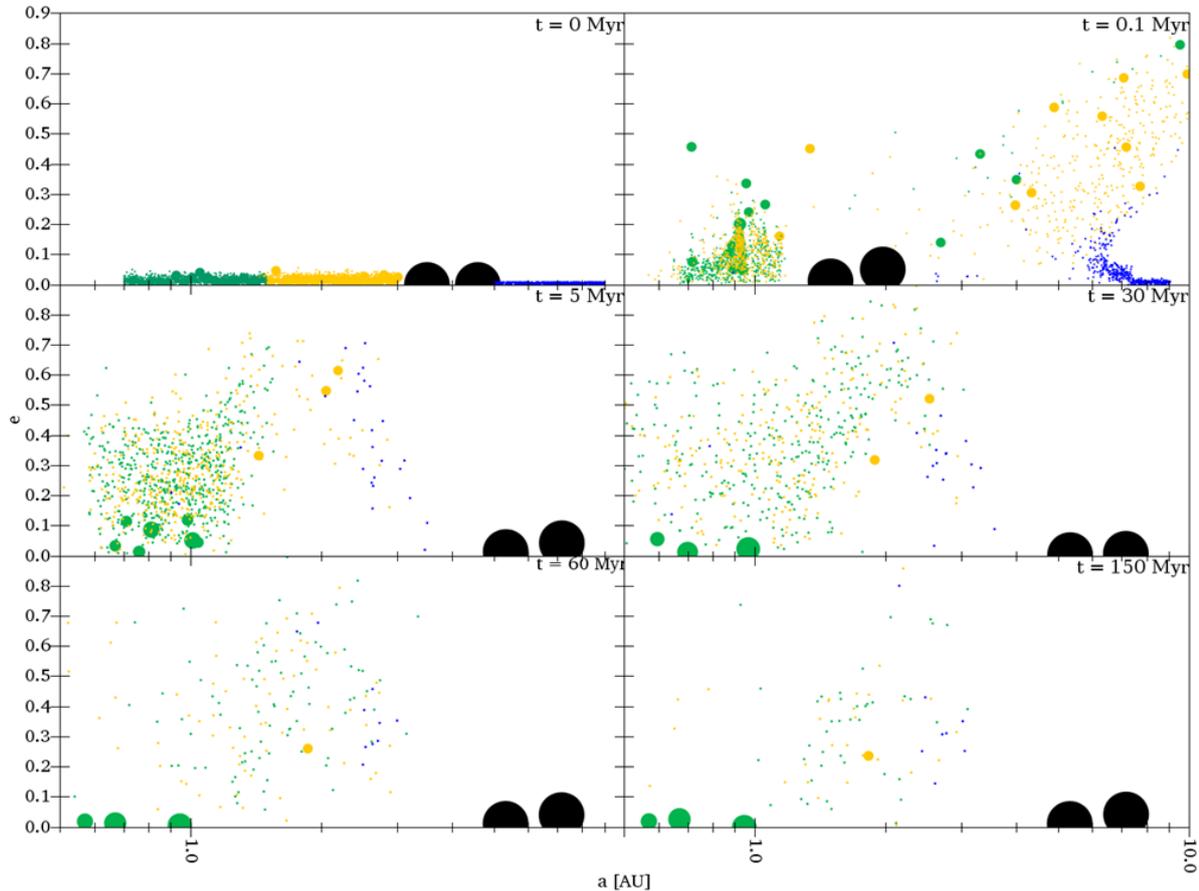

**Fig. 2:** Snapshots of the evolution of a Grand Tack simulation. The colour coding is green for bodies that began closer than 1.5 AU to the Sun, orange for bodies originally between 1.5 AU and 3.0 AU and blue for material beyond that. Jupiter and Saturn are depicted as large black bullets. The symbol sizes are proportional to the mass of the bodies; for Jupiter and Saturn a fixed size was chosen so as not to fill most of the panels.

After 5 Myr of simulation evolution the gas giants have reversed their migration and stalled close to their current positions (bottom-left). They have emptied out the asteroid belt region and left behind predominantly EC material in the inner Solar System. Earth and Venus have reached approximately half of their masses. By 30 Myr (middle-right) the terrestrial planets are almost fully formed and not much happens after 60 Myr of simulation evolution (bottom-left) to 150 Myr (bottom-right).

The Classical model and the Grand Tack model predict slightly different outcomes for the bulk compositions of the terrestrial planets (Woo et al., 2018). In the Classical model, the planets tend to accrete mostly locally. In contrast, in the Grand Tack model, all the accretion is confined to an annulus between approximately 0.5 AU and 1 AU, but this region also contains material that originated in the outer parts of the protoplanetary disk that was scattered inwards by the migration of Jupiter. Additional contamination of the inner Solar System with material from the outer Solar System occurs during the growth of the gas giants (Raymond and Izidoro, 2017b).

## 2.3 Alternative model hypotheses

We briefly mention several alternative models of terrestrial planet formation, which have either not (yet) gained much traction or are too new to have been thoroughly explored, especially for their cosmochemical implications. The first is the annulus model (Agnor et al., 1999; Hansen, 2009; Walsh and Levison, 2016), wherein solid material in the disk is confined to a narrow region around the Sun, typically between 0.5 AU or 0.7 AU to 1 AU or 1.5 AU. The justification for the inner edge of the annulus model is due to the large mass ratio between Mercury and Venus; in a similar vein, the justification for the outer edge of the annulus is to attempt to reproduce the large mass ratio between Earth and Mars. The rationale for the edges is that a low density of solid material will hinder the Mercurian and Martian embryos from gathering it so that they cannot grow into fully fledged planets such as Venus and Earth.

The second is the depleted disk model (Izidoro et al., 2014; 2015) wherein the mass of the asteroid belt is initially much lower than elsewhere in the protoplanetary disk; a variation of this model (Raymond and Izidoro, 2017a) fills the asteroid belt region up with the outwards scattering of material from the terrestrial planet formation zone and the inwards scattering of material from the giant planet region during giant planet migration. This variation is, in essence, no different from the annulus model. In this model, Mars' small mass compared to Earth's is caused by the density drop of solid material around 1 to 1.5 AU. As such, the Mars analogue was unable to accrete much material once the Martian embryo had already mostly grown by locally accreting material. The reason for the density drop is left unspoken.

A third model (Clement et al., 2019) invokes a dynamical instability in the outer Solar System wherein the giant planets scatter each other and all material around them. There is ample evidence that such a dynamical instability has occurred (Thommes et al., 1999; Tsiganis et al., 2005; Gomes et al., 2005; Morbidelli et al., 2005; Brasser et al., 2009; Morbidelli et al., 2010; Levison et al., 2011; Nesvorný and Morbidelli, 2012; Nesvorný, 2015a, 2015b; Brasser and Lee, 2015), but thus far it has been assumed that this instability happened after terrestrial planet formation to account for the late heavy bombardment purported to be recorded by the formation of the large lunar impact basins (Tera et al., 1974). In Clement et al. (2019) the authors assume that this dynamical instability occurred early, while the terrestrial planets were still forming. Mars' low mass is then a result of material in its vicinity being removed by perturbations from the migrating giant planets as they evolved through a high-eccentricity phase. Based on evidence from reset ages of meteorites for radiometric systems with different closure temperature Mojzsis et al. (2019) showed that giant planet migration must have commenced before 4480 million years ago (Ma), lending some support to the early instability model of Clement et al. (2019).

A fourth variation is the pebble accretion model (Ormel and Klahr, 2010; Lambrechts and Johansen, 2012; Johansen and Lambrechts, 2017) wherein Sun-wards-drifting pebbles accrete onto large planetesimals or small planetary embryos (Johansen et al. 2015). The high growth rate of pebbles compared to planetesimals is due to dissipation of kinetic energy by gas drag as a pebble is scattered by a growing protoplanet. These embryos then rapidly grow into either the giant planets (Levison et al., 2015a) or the terrestrial planets (Levison et al., 2015b). The radial pebble flux through the protoplanetary disc is a key parameter that determines the final masses of the embryos. Lambrechts et al. (2019) demonstrated that low pebble fluxes lead to the formation of Mars-mass embryos that then go on to collide as in the classical terrestrial planet formation picture. Larger pebble fluxes lead instead to the formation of super-Earths with masses between 1 and 10 times the mass of the Earth. The pebble accretion model is actively studied both in the Solar System and for exoplanets (e.g. Johansen and Lambrechts, 2017; Matsumura et al., 2017; Izidoro et al., 2019). Furthermore, this model also explores the idea of drifting pebbles piling up in the terrestrial planet region, first forming planetesimals and subsequently the terrestrial planets (Drazkowska et al., 2016). This model is then a precursor for the annulus or depleted disk models.

### 2.4 The terrestrial planet formation chart

A good model of terrestrial planet formation should match a variety of criteria. The criteria that we consider here are: fast accretion of mass ($< 30$ Myr; e.g., Kleine et al., 2002; Yin et al., 2002; Yu and Jacobsen, 2011; Kleine and Walker, 2017, and references therein), the mass-semimajor axis (M-a relation) distribution of the planets, the low mass of Mars (closely related to the first criterion), the angular momentum deficit (AMD) of the planets, the radial concentration distribution ($S_r$), the timing of the Moon-forming event (Lock et al., 2020, this issue), the isotopic composition of the planets, and the lunar and Martian cratering chronology. We have composed a chart wherein we list each of the criteria from relevant literature sources and whether the models listed above match the criteria. When certain criteria have not yet been tested, we write 'U' for 'Unknown'. Table 1 is filled with the results from the studies listed in the previous subsections, as well as those of O'Brien et al. (2006; 2014), Jacobson et al. (2014), Brasser et al. (2018, 2020) and Woo et al. (2018). The criteria for success vary from one publication to another. Here we have taken the results from literature sources and tallied the successes or failures based on what the authors wrote in their respective publications. Where unclear, we e-mailed the authors with specific requests. An exhaustive list of individual successes and failures is beyond the scope of this work.

**Table 1:** Terrestrial planet formation comparison chart related to the different accretion stages I, II, III, and IV illustrated in Fig. 1. The criteria and the symbols are explained in the main text.

| Accretion stages | I-III | IV | | | | | | |
|---|---|---|---|---|---|---|---|---|
| Model | Fast mass accretion | M-a relation | Mars | AMD | $S_T$ | Moon | Composition | Chronology |
| Classical | ✗ | ✗ | ✗ | ~ | ✗ | ✓ | ✓ | U |
| Annulus | ✗ | ✓ | ✓ | ~ | ✓ | ✓ | U | U |
| Grand Tack | ✗ | ✓ | ✓ | ✓ | ✓ | ✓ | ✓ | ✓ |
| Depleted disk | ✗ | ✓ | ✓ | ✓ | ✗ | U | U | U |
| Pebble | ✓ | ~ | ~ | U | U | U | U | U |
| Early instability | ✓ | ~ | ~ | ~ | ✗ | ✓ | U | U |

We consider the model to work if the success probability for a particular event exceeds 50 % in the relevant literature sources (symbol ✓), inconclusive if it falls from 25 % to 50 % (symbol ~) and a failure if it is below 25 % (symbol ✗). *We caution against accepting these results at face value because different studies have different criteria for what is considered a successful semi-major axis relationship, or a mass threshold for a Mars analogue.* This table should be considered as *indicative only* and prompt further investigation into particular aspects of each model.

### 3. Terrestrial planet accretion rates constrained by isotopes

As briefly reviewed above, planetesimals and planetary embryos began to form from solid grains originating from the gas and dust in the disk ≈ 4567 Ma (e.g.; Jacobsen, 2003; Bollard et al. 2017). Planetesimals then formed within the first few $10^5$ years (Kruijer et al., 2014) followed by planetary embryos, such as Mars, over the next few million years (Dauphas & Pourmand, 2011). The terrestrial planets subsequently formed from accretion of Moon- and Mars-sized planetary embryos that originated at various heliocentric locations. During the past decades, the terrestrial planet formation models and scenarios discussed above used different accretion rates, timing of metal-silicate partitioning, and compositions of planetary building blocks. However, the before-mentioned different model approaches and parameter spaces that are applied by numerous planetary formation modellers (e.g., Morbidelli et al., 2012; Raymond et al., 2004; 2009; O`Brien et al., 2006; Johansen et al., 2007; Izidoro and Raymond, 2018; Raymond and Morbidelli, 2020; and references therein), and core formation models (magma ocean model, two-stage model, local equilibrium model, see: Jacobsen et al., 2008; and references therein) yield different time scales for the formation of large terrestrial planets like Earth or Venus that lie within a range of ≤ 30 - 100 Myr.

Since measurable variations in isotopic ratios of certain elements in terrestrial planets and their atmospheres, as well as in meteorites, provide records of accretion, formation of the core and crust, we briefly review the so-called isotopic chronometers $^{182}$Hf-$^{182}$W and U-Pb (e.g., Jacobsen and Harper, 1996; Yin et al., 2002; Halliday, 2004; Kleine et al., 2004; Rudge et al., 2010; Jacobsen et al., 2008; Yu and Jacobsen, 2011; Kleine and Walker, 2017), and isotopic systems such as $^{50}$Ti, $^{54}$Cr, $^{92}$Mo and $^{100}$Ru (e.g., Dauphas, 2017; Brasser et al., 2018; Woo et al., 2018; Carlson et al., 2018), $^{48}$Ca/$^{44}$Ca isotope systematics (Schiller et al., 2018), atmospheric $^{36}$Ar/$^{38}$Ar, $^{20}$Ne/$^{22}$Ne, $^{36}$Ar/$^{22}$Ne ratios (Marty, 2012; Lammer et al., 2020b), the evidence of solar $^{3}$He abundance in Earth's deep mantle and the D/H ratio in Earth's sea water, that can be used to constrain the accretion time scales of the terrestrial planets.

### 3.1 Radiometric dating systems

#### 3.1.1 $^{182}$Hf-$^{182}$W chronometry for Earth and Mars' accretion

The fractionation of gas and dust during the stages of disk formation and accretion of planetesimals in the solar nebula (< 5 Myr, Bollard et al. 2017) led to a fractionation between volatile and refractory elements (e.g., Wai and Wasson, 1977; Jacobsen et al., 2008). At this early time, Hafnium (Hf) and Tungsten (W) isotopes are not significantly fractionated. The differentiation of planetesimals, planetary embryos and protoplanets, however, leads due to the separation of metal-sulphide melt from silicates to the formation of cores resulting in Hf-W fractionation (e.g., Jacobsen et al., 2008; Kleine and Walker, 2017). The Hf-W-system therefore provides the most reliable radiometric chronometer of core formation for terrestrial planets such as the Earth (e.g., Jacobsen and Harper, 1996; Harper and Jacobsen, 1996; Lee and Agee, 1996; Halliday and Lee, 1999; Halliday and Lee, 1999; Halliday 2000; Kleine et al., 2002; 2004; Yin et al., 2002; Jacobsen et al., 2005; Wood and Halliday, 2005; Jacobsen et al., 2008; Kleine et al., 2009; Yu and Jacobsen, 2011; Kleine and Walker, 2017).

The elements Hf and W are both highly refractory trace elements. Hafnium is a lithophile element and strongly partitioned into the silicate portion of a planet, while W is moderately siderophile and partitioned preferentially into the coexisting metallic phase. Therefore, ≥ 90 % of Earth's W has moved into the core during its formation (Halliday, 2000) (see Fig. 3a), making the history of planetary bodies and the timing of accretion, differentiation, and core formation a valuable dating technique in planet formation (Kleine et al., 2009).

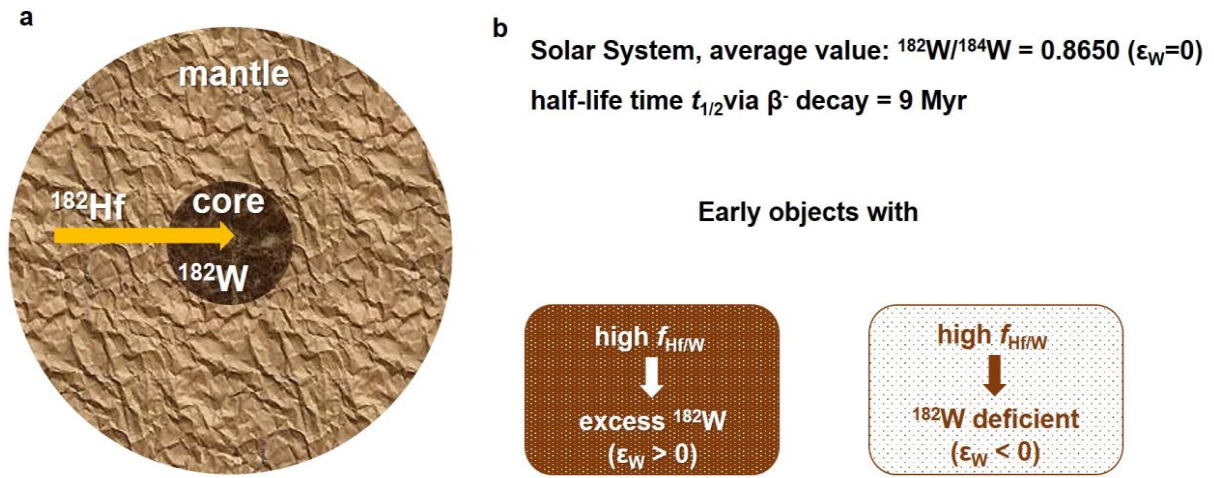

**Fig. 3**: Panel a: The lithophile element 182Hf remains in the mantle while the moderately siderophile element $^{182}$W moves preferentially into the metallic core. Because $^{182}$Hf decays with a half-life time of only 9 Myr via $^{182}$Ta into $^{182}$W these elements represent a precision timer of accretion, differentiation, and core formation of terrestrial planets. Panel b: Illustration of the essential Hf-W feature (after Halliday, 2000).

The basic principles of the $^{182}$Hf-$^{182}$W chronometry method are briefly described below. $^{182}$Hf decays to $^{182}$W via $^{182}$Ta with a short half-life time of 8.9 Myr. Initially, their parent and trace elements should be in chondritic portions ≈ 1 : 1 in the early Solar System bodies because both elements are highly refractory (e.g., Jacobsen and Harper, 1996; Halliday and Lee, 1999; Halliday 2000; Kleine and Walker, 2017).

As mentioned above Hf is lithophile and remains in the mantle but W, which exists preferentially into the metallic phase, sinks into the core. Therefore, the parent-to-daughter ratio $f_{Hf/W}$ is perturbed by the formation of the core (e.g. Kleine and Walker, 2017; and references therein). The residual silicate portion in Earth's crust and mantle is commonly described as the bulk silicate Earth (BSE) and has a $f_{Hf/W}$ in the range of ≈10-40 (Newsom et al., 1996). More recent studies that correlated W with other refractory lithophile trace elements such as U, Th, Ta and Ba (McDonough, 2003; Dauphas and Purmand, 2011; König et al., 2011) indicate that $f_{Hf/W}$ of the BSE is 18.8 ± 5 (Kleine and Walker, 2017). This value is an order of magnitude higher than the chondritic value (see Fig. 3b). If the fractionation of Hf from W caused by core formation takes place during the lifetime of $^{182}$Hf, an excess $^{182}$W relative to other isotopes of W should develop in the silicate portion of the planet because of an enhanced $f_{Hf/W}$ as illustrated in Fig. 3b.

By using $^{182}$Hf-$^{182}$W chronometry and the ratio $f_{Hf/W}$ one can now constrain the rate of Earth's accretion and the timing of core formation by geochemical models. Several Moon- to Mars-mass planetary embryo impactors released enough energy that proto-Earth produced temporary magma oceans so that core formation could take place (e.g., Rubie et al., 2007).

The segregation of metal and hence core formation occurs within the early magma ocean state with a metal-silicate equilibrium at very high pressure and temperature (e.g., Rubie et al., 2003; Jacobsen et al. 2003; Kleine and Walker, 2017). Core formation models assume that the terrestrial planets accrete dust and planetesimals with a chondritic Hf/W ratio related to an Hf and W-isotopic composition. Jacobsen and Harper (1996) suggested a model that assumes local equilibrium at which the newly added material to the growing protoplanet immediately differentiates into metal and silicate, with the metal sinking to the core without equilibrating with the remaining silicate mantle. However, this local equilibrium model results in a very long mean-time for Earth's core formation of ≈ 108 Myr (≈ 4460 Ma) with a 90 % core-formation time of the order of ≈ 250 Myr (≈ 4318 Ma) (Jacobsen and Harper, 1996; Jacobsen et al., 2008). A reason for this long time-scale is that the newly segregated core material in this model reflects only the average isotopic composition of the delivered material.

This long core formation mean-time is also not in agreement with the strong evidence for an early silicate differentiation at ≈ 4530 Myr ago from measured $^{143}$Nd/$^{144}$Nd and $^{142}$Nd/$^{144}$Nd ratios of the 3800 Myr ago supracrustal rocks of Isua from West Greenland (Harper and Jacobsen, 1992; Jacobsen and Harper, 1996; Boyet et al., 2003; Caro et al., 2003; Boyet and Carlson, 2005; Bennet et al., 2007; Jacobsen et al., 2008). These findings let to a general consensus that the related Sm-Nd-systematics as fractionation of Sm/Nd taking place within the first 100 Myr after the origin of the Solar System, i.e. before 4468 Ma. This would also indicate that the Moon-forming event, which led to the last magma ocean, occurred within the first 100 Myr (e.g. Touboul et al., 2007; Asphaug, 2014; Lock et al., 2020, this issue).

Lee and Agee (1996) developed a so-called metal pond core formation model, where the authors assumed that the newly segregated material is equilibrated at the bottom of the magma ocean at roughly 1000 km. Harper and Jacobsen (1996) and Jacobsen (2005) developed the so-called magma ocean differentiation model, where the newly segregated core material reflects the average isotopic composition of the silicate mantle due to full equilibrium in a magma ocean. Today, the magma ocean model is the most widely used core formation model related to the interpretation of early Earth's Hf-W-chronometry studies (e.g., Yin et al., 2002; Wood and Halliday, 2005; Jacobsen et al., 2008; Kleine et al., 2009; Yu and Jacobsen, 2011; Kleine and Walker, 2017). Two-stage accretion scenarios yield results between the magma ocean model with complete equilibration between metal and silicates and the local equilibrium model (e.g., Jacobsen et al. 2008; references therein).

Experimental work and model results indicate that the primitive mantle abundances of the moderately siderophile elements Fe, Ni, Co, Mo, W, P, the moderately siderophile element Ge and the highly siderophile element Re are consistent with a so-called metal-silicate equilibrium at the base of a 700 to 1000 km deep magma ocean in the early Earth, and hence a homogeneous accretion (Righter and Drake, 2000; Righter et al., 2011). During the stage of planetary embryo formation surface temperatures can reach up to 2000 - 3000 K (Ikoma and Genda, 2006; Bouhifd and Jephcoat, 2011; Stökl et al., 2016), so that deep magma oceans would have been present (Lammer et al., 2020a, this issue; 2020b).

By assuming complete metal-silicate equilibrium the Hf-W systematics yield a fast accretion of the main mass of Earth within 10 Myr (Wade and Wood, 2005; Cogne et al., 2008; Jacobsen et al., 2008). The Moon-forming event, however, finished Earth's main phase of accretion between ≥30 Myr (e.g., Jacobsen, 2005; Albarède, 2009; Barboni et al., 2017; Thiemens et al., 2019) to ≈ 100 Myr (e.g., Halliday, 2008; Halliday and Wood, 2009; Boyet and Carlson, 2005; Connelly et al. 2019) i.e. between 4538 Myr and 4468 Myr ago. Metal segregation happens faster than accretion (Jacobsen and Harper, 1996; Jacobsen et al., 2008, and references therein), so that the time scale of core formation is generally used to constrain Earth's accretion rate (e.g., Rudge et al., 2010; Yu and Jacobsen, 2011; Kleine and Walker, 2017; and references therein). Equilibrium two-stage model results for Hf-W systematics agree with accretion ages of ≈ 31.0 ± 4.4 Myr (Rudge et al., 2010; Kleine and Walker, 2017), meaning that core formation was established by 4537 ± 4.4 Myr ago.

According to these studies, the $^{182}$Hf-$^{182}$W chronometry method indicates that at least 80 % of the early Earth must have accreted in ≤ 35 Myr, while the final ≈10 % must have been accreted by 120 Myr. In partial metal-silicate equilibrium models, proto-Earth would have accreted ≈ 0.63 $M_{Earth}$ after a duration of some 10 - 25 Myr (Jacobsen et al., 2008; Mamajek, 2009; Rudge et al., 2010; Kleine and Walker, 2017; Morbidelli, 2018).

Yu and Jacobsen (2011) showed that one could overcome discrepancies mainly explained by metal-silicate disequilibrium during giant impacts in Hf-W systematics by including constraints from partitioning of refractory siderophile elements Ni, Co, W, V, and Nb during core formation. These authors found that the concentrations of the siderophile elements in the mantle are consistent with high-pressure metal-silicate equilibration in a terrestrial magma ocean (Righter and Drake, 2000; Righter et al., 2011). In this study, the earliest time the Moon-forming event could have taken place is at roughly 30 Myr after the origin of the Solar System.

Under these assumptions, the model of Yu and Jacobsen (2011) indicate that the timing of the Moon-forming event is inversely correlated with the time scale of Earth's main accretion stage of the first 0.87 $M_{Earth}$.

If the Moon-forming giant impact occurred later than 50 Myr (i.e. after 4510 Myr ago), the main stage of early Earth's accretion would need to have occurred before < 10.7 ± 2.5 Myr, while an earlier Moon-forming event would increase the time taken for Earth's primary accretion (Yu and Jacobsen, 2011). To summarise, the evidence for Earth's (terrestrial planet) accretion from $^{182}$Hf-$^{182}$W chronometry within its uncertainties results in a time scale for the main mass accretion of 8.2 - 35 Myr and proto-Earth masses at the time when the disk dissipated at 3.5 – 4 Myr (Bollard et al., 2017; Wang et al., 2017) of 0.5 – 0.75 $M_{Earth}$, if one assumes full metal-silicate equilibrium (Yin et al., 2002; Jacobsen et al., 2008; Rudge et al., 2010; Yu and Jacobsen, 2011; Kleine and Walker 2017). Yu and Jacobsen (2011) obtain a mass for proto-Earth at roughly 4 Myr that is of the order of 0.56 $M_{Earth}$ with a Hf/W ratio of ≈ 15 and a late giant Moon-forming impact ≥ 52 Myr (i.e. after 4510 Myr ago).

As was shown in Lammer et al. (2020a, this issue; 2020b), if proto-Earth accreted in excess of > 0.58 $M_{Earth}$ during the lifetime of the solar nebula, the present-day atmospheric $^{36}$Ar/$^{38}$Ar, $^{20}$Ne/$^{22}$Ne and $^{36}$Ar/$^{22}$Ne ratios (see also Sect. 3.1.4) cannot be reproduced through EUV-driven hydrodynamic escape of a thin primordial atmosphere. Moreover, these authors also show that if the proto-Earth accreted > 0.7 $M_{Earth}$ during the disk lifetime (≈ 4 Myr), it would need hundreds of Myr or even Gyrs to lose its primordial atmosphere. One can therefore conclude that the upper mass of proto-Earth at 4.0 Myr, i.e. the time when the disk dissipated, should have been ≤ 0.58 $M_{Earth}$ (Lammer et al., 2020b).

The analysis of $^{182}$Hf-$^{182}$W chronometry for dating Mars' accretion is more complex because one cannot measure the Hf/W ratio of bulk silicate Mars (Kleine and Walker, 2017). In Mars' case one must compare the abundance of W to other lithophile elements that have the same incompatibility as W. If one assumes that refractory elements such as U, Th, Hf and W occur in chondritic relative portions in bulk Mars, than Hf/W of bulk silicate Mars can be estimated from U/W × chondritic Hf/U or Th/W × chondritic Hf/Th of Martian rock samples (Dauphas and Pourmand, 2011). By using this assumption, two-stage accretion models yield a core formation age for Mars of 4.1±2.7 Myr, while exponential growth models for Mars yield ≈ 63 % accreted mass at roughly 2.4 Myr and up to ≤ 10 Myr for the remainder of Mars' accretion (Dauphas and Pourmand, 2011; Kleine and Walker, 2017). Using $^{60}$Fe-$^{60}$Ni chronometry, Tang and Dauphas (2014) independently obtained the same accretion time scale. Thus, one can say that Mars was almost fully formed when the disk dispersed after 3.3 – 4.5 Myr (Bollard et al., 2017; Wang et al., 2017).

### 3.1.2 Uranium-lead (U-Pb) isotopic dating system

The U-Pb dating system is one of the earliest radiometric dating schemes that was used for dating rocks that crystallised between millions of years to over 4500 Ma (e.g., McDonough and Sun, 1995; Kellogg et al., 2007; Jacobsen et al., 2008; Schaltegger et al., 2015). Because the exact rate at which U decays into Pb is well known, the current ratio of Pb to U in a rock sample can be used to determine its age, provided that in the interim no Pb loss occurred. The U-Pb dating system relies on two separate isotope decay chains: the first one is from $^{238}$U to $^{206}$Pb, with a half life of 4.47 Gyr, while the second decay chain begins at $^{235}$U which decays to $^{207}$Pb with a half life of 710 Myr. U-Pb systematics applied to Earth's mantle suggest that the time it takes for the Earth to grow to ≈ 0.63 $M_{Earth}$ is 56 – 130 Myr for an equilibrium two-stage model, and 21.5 – 51 Myr for an exponential growth model (Rudge et al., 2010).

Compared to $^{182}$Hf-$^{182}$W systematics of Earth's mantle discussed above, U-Pb systematics yield accretion time scales that are much longer (e.g., Halliday, 2004; Kleine et al., 2004; Rudge et al., 2010; Kleine and Walker, 2017). Because U-Pb chronometers are long-lived chronometers that are susceptible to geological processes throughout Earth's history, as well as at its beginning, this method does not give as strong an accretion constraint as the Hf-W chronometer (Jacobsen et al., 2008; Kleine and Walker, 2017; and references therein).

The blanketing effect of the $H_2$/He-dominated primordial atmosphere that surrounded growing protoplanets during and, depending on their mass, also some time after the disk lifetime, as well as collisions between Moon- to Mars-mass planetary embryos that were involved in the growth of proto-Earth and Venus, released enough energy to produce a global magma ocean (e.g., Jacobsen et al., 2008; Lammer et al., 2020a, this issue; 2020b), so core formation as discussed before started much earlier than estimated by the U-Pb chronology (e.g., Rubie et al., 2007; Kleine and Walker, 2017). Due to this discrepancy, Kamper and Kramers (2006) and Yin and Jacobsen (2006) suggested that U-Pb systematics do not constrain core formation since bulk Earth's U/Pb ratio and the Pb isotopic composition are not known well enough. Moreover, the discrepancy in accretion time scales related to U-Pb systematics may be caused by a late segregation of Pb-bearing sulphides into Earth's core (Wood and Halliday, 2005; Halliday and Wood, 2007), disequilibrium during core formation (Halliday, 2004; Allègre et al., 2008), and the possible addition of Pb during the late veneer (Albarède, 2009). To summarise this brief discussion: according to Rudge et al. (2010) and Kleine and Walker (2017), one can conclude that U-Pb systems contain little information on the early accretion, but constrain Earth's late accretion phase very well.

However, the main uncertainties between Hf-W and U-Pb systematics for the determination of the core formation time-scale is the strong dependence on the metal-silicate equilibrium (e.g., Kleine et al., 2002; Allègre et al., 2008) and the uncertainties in knowledge related to the bulk Earth Pb isotopic composition and U/Pb ratio (e.g., Yin and Jacobsen, 2006). Rudge et al. (2010) showed that Hf-W constrains the early phase of accretion, while the U-Pb system constrains the late stage of accretion but contains little information on Earth's early growth. One can bring the Hf-W and U-Pb age determinations into agreement if one assumes a full metal-silicate equilibrium combined with a very fast early accretion that subsequently slowed down and was followed by a long phase of much slower growth culminating in a late giant impact, i.e., the Moon-forming event (Rudge et al., 2010; Yu and Jacobsen, 2011).

Moreover, Kellogg et al. (2007) showed that a mean age of accretion and core formation of $\approx$ 10 Myr is consistent with the isotopic composition of modern mid-ocean ridge basalts when taking into account recycling of continental and oceanic crusts into the mantle over time. This result agrees with the study of Rudge et al. (2010) that there is no need to infer a late core formation to explain Earth's Pb isotopic composition. According to Jacobsen et al. (2008) it seems that the fractionation of U-Pb in the nebula is, a dominant process so that the core-related fractionation has to be very accurately determined compared with the volatile-related fractionation in order to obtain accurate results for the timing of core formation based on the U-Pb-system. Therefore, the U-Pd-isotopic dating system does not place strong constraints on the chronology of accretion and core formation.

## 3.2 Constrains from lithophile-siderophile isotope data, $^{48}Ca/^{44}Ca$-systematics and noble gases

### 3.2.1 Earth's accretion constrained by lithophile-siderophile elements

For decades, researchers have tried to estimate Earth's bulk chemical composition by comparison with meteorites. Clayton et al. (1970) found that Earth's oxygen composition is different compared to most meteorites apart from the silicon-rich and highly reduced enstatite chondrites. However, the chemical composition of enstatite chondrites is very different compared to rocks on Earth's surface (e.g., Drake and Righter, 2002; Carlson 2017), so the community continued to assume that the more oxidised and volatile-rich carbonaceous chondrites were involved in early Earth's accretion.

Recently Dauphas (2017) used the unique isotope content of different types of meteorites to identify those that could have been involved in Earth's accretion over time. He calculated that the isotopic mantle signatures of lithophile elements such as O, Ca, Ti and Nd, and moderate to highly siderophile elements like Cr, Ni, Mo and Ru recorded different accretion stages of the early Earth. Dauphas (2017) showed that these elements point to a large fraction of Earth's building blocks, some 70%, that were isotopically similar to enstatite meteorites. It was found that for the first 60% of its accretion, proto-Earth most likely formed from a mixture of an isotopic composition of 51% enstatite, 40 % ordinary chondrites and 9 % carbonaceous chondrites (i.e. 5.4% CC in total). For the second accretion stage, i.e. the next 60 – 99.5% of Earth's mass, and for the final 0.5% accreted during the late veneer, the isotopic record of these elements are mainly akin to enstatite chondrite-like material.

At the same time, Fischer-Gödde and Kleine (2017) concluded from Ru isotopic measurements from a sample of meteorites that material arriving as late accretion was relatively dry and isotopically identical to enstatite chondrites. This finding is also in support of Dauphas (2017) that states enstatite chondrites most likely played an important role in the accretion of the Solar System's terrestrial planets, although in an earlier study Zhang & Becker (2013) state that S, Se and Te abundance in the terrestrial mantle requires a volatile-rich late veneer. Carlson et al (2018) conclude that the composition of the planetary building blocks changes over time and Earth's deep interior should have a different composition compared to its surface layers. However, the findings and three-stage accretion model presented by Dauphas (2017) neither gives a time scale for the two main accretion stages nor describes how much mass was accreted while the planet was embedded within the disk or after its dispersal.

3.2.2 Accretion rates of inner disk mass versus time obtained from $^{48}Ca/^{44}Ca$ isotope data

Current models that study the formation of planetary embryos from dust settling to planetesimal formation via pebbles (that originate from the same flow) struggle to explain the different compositions of Solar System bodies (e.g., Johansen et al., 2015; Morbidelli, 2018). Schiller et al. (2018) measured the $^{48}Ca/^{44}Ca$ isotope content in samples from Earth, Mars, Vesta, and the angrite and ureilite parent bodies. Here, $\mu^{48}Ca$ corresponds to the $^{48}Ca/^{44}Ca$ ratio relative to that in the terrestrial reference standard, given in parts per million (p.p.m.). A positive correlation between $^{48}Ca/^{44}Ca$ isotope ratios and planetary masses was identified. To explain these findings, a highly debated and unorthodox view of planetary growth is proposed by these authors. Schiller et al. (2018) suggest that pebble accretion is correct, however, the different bodies were growing at the same rate but stopped growing at different times (Schiller et al., 2018; Morbidelli, 2018). This hypothesis is non-conventional because the mainstream theory is that bodies of different masses and at different locations grew at different rates during the lifetime of the disk. Fig. 4 illustrates their hypothesis of planetary accretion. In the model approach of Schiller et al. (2018), the gas and dust in the disk surrounding the young Sun initially contained inner-disk material with low $\mu^{48}$Ca ≈ -150 ppm, similar to the ureilites, and outer disk particles with high $\mu^{48}$Ca values of ≈ +200 ppm, which corresponds to carbonaceous chondrites.

In this scenario, the planetesimals and planetary embryos grew by accreting matter from the inner disk. The inner disk started to be fed with inwards-drifting pebbles and, after some time, the inner-disk material was largely supplemented by the material from the outer disk, resulting in an increase of the mean $\mu^{48}$Ca value of the inner disk. The planetary embryos continued to grow until they were displaced by perturbations from other bodies. The assumed $\mu^{48}$Ca mixing from the inner and outer material resulted in values of $\mu^{48}$Ca ≈ -100 p.p.m. for Vesta, ≈ -20 p.p.m for Mars and 0 p.p.m. for Earth and the Moon. From their analysis, the authors conclude that proto-Earth grew to ≈ 0.8 $M_{Earth}$ within the disk lifetime and its accreted material consisted of ≈ 60 % Ureilite and ≈ 40% CI-chondrite-like material (which are the end-members of the model). This rapid accretion is difficult to reconcile with current dynamical and Hf-W chronometry models (e.g. Carlson et al., 2018; Yu and Jacobson, 2011).

In a brief report, Morbidelli (2018) remarked various inconsistencies within the hypothesis proposed by Schiller et al. (2018). One problem is that, for the Earth, the model of Schiller et al. (2018) requires about 0.4 $M_{Earth}$ to be accreted from the outer disk, which is higher than previous calculations by Dauphas (2017) and Morbidelli et al. (2016). However, the previously mentioned studies did not include ureilite-like material in their models because of the inability from the existing samples of the Ureilites and other achondrites to obtain the bulk elemental concentrations of the parent body that the models require.

The added outer Solar System material is represented isotopically in Ca ratios of CI chondrites (Schiller et al., 2018). The composition of the proto-Earth as inferred by Schiller et al. (2018) is not in agreement with the isotope measurements of the lithophile, moderately and highly siderophile elements of the material out of which Earth's mantle accreted, as determined by Dauphas (2017), since it does not include a high amount of enstatite material (see also Sect. 3.1.2) because Schiller et al. (2018) focused on an end-member model only. Furthermore, it is unlikely that a terrestrial mantle consisting of 40% CI chondrite is consistent with the terrestrial Mg/Si elemental ratio as well as with oxygen, titanium, chromium and other isotopic systems. Recently, Schiller et al. (2020) claim that terrestrial Fe isotopes are nearly identical to those of CI chondrites. From their measurements they imply that the terrestrial mantle consists of 40% CI. It is clear that further studies are needed to resolve this dilemma.

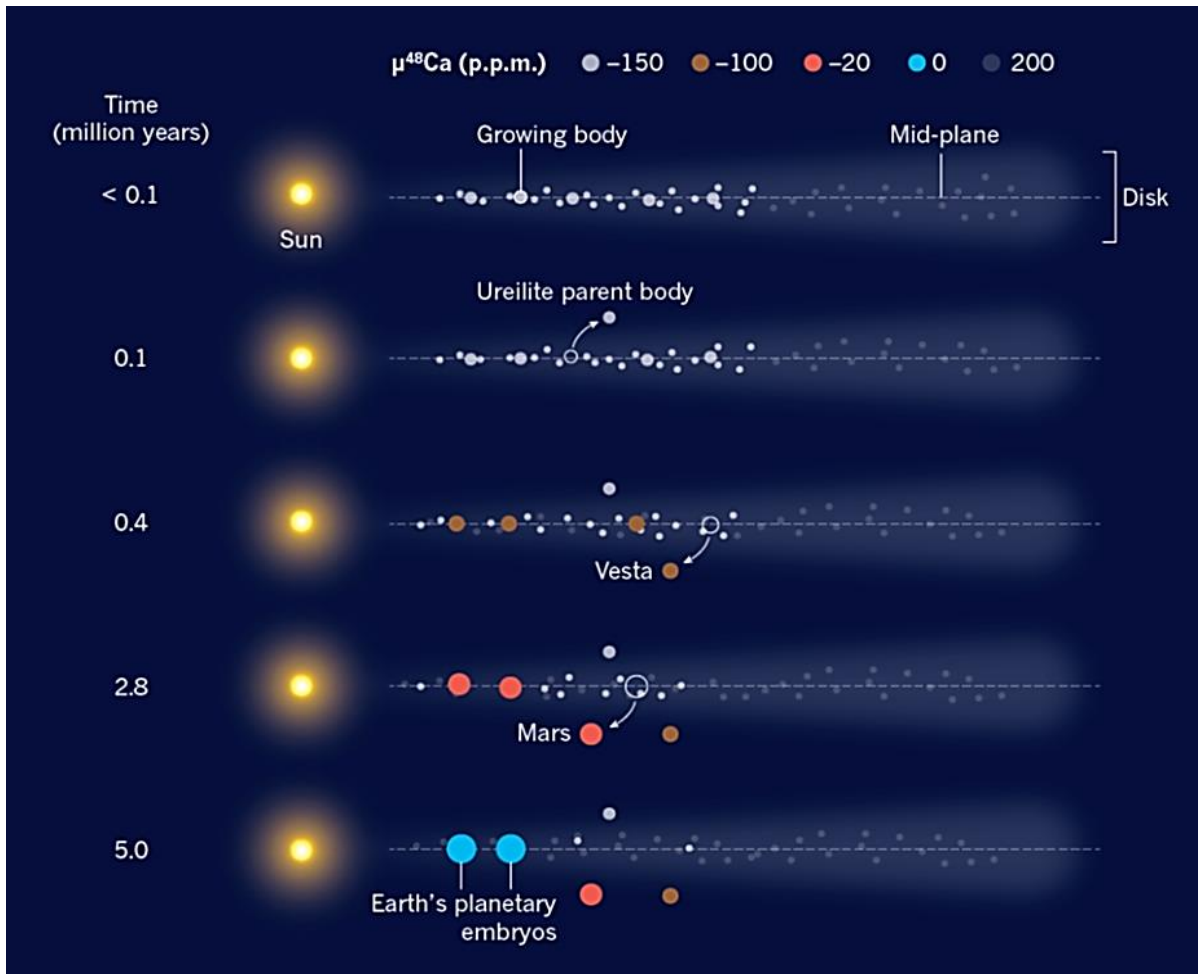

**Fig. 4**: Illustration of the hypothesis suggested by Schiller et al. (2018) for the accretion of Venus, Earth and Mars based on measured $^{48}Ca/^{44}Ca$ isotope ratios ($\mu$ $^{48}Ca$ [p.p.m.]) that correlate between the masses of the inner Solar System bodies. The times indicate accretion stages during the disk's lifetime in which various bodies originated. In this scenario, the parent bodies of Ureilite-type meteorites, the asteroid Vesta, Mars and planetary embryos that probably formed Earth are also shown (from Morbidelli, 2018).

From atmospheric $^{20}Ne/^{22}Ne$, $^{36}Ar/^{38}Ar$ and $^{36}Ar/^{22}Ne$ reproduction attempts, Lammer et al. (2020b) concluded that protoplanets that accreted masses of $> 0.58$ $M_{Earth}$ within the disk lifetime of 3.3 – 4.5 Myr (Bollard et al., 2017; Wang et al., 2017) would capture a primordial atmosphere that is too massive to be consistent with the present atmospheric ratios, while a proto-Earth with a mass of $\geq 0.7$ $M_{Earth}$ at 4 Myr likely ends up as a gas dwarf (e.g., Owen and Wu, 2016; Fossati et al., 2017; see also Lammer et al. 2020a; this issue).

As shown in Fig. 5, it is possible to reproduce the present-day atmospheric Ar and Ne noble gas ratios of the Earth with a composition as suggested by Schiller et al. (2018) but not by growing up to 80% of its mass within the nebula (that dispersed at 3.3 – 4.5 Myr). In such a case proto-Earth accretes up to $\approx 0.53 – 0.58$ $M_{Earth}$ within the gas disk and grows fast up to $\geq$ 0.8 $M_{Earth}$ immediately afterwards and still before the escape of a tiny captured primordial atmosphere (within < 7 Myr) through planetesimal accretion. Afterwards mainly CCs or CIs – as suggested by Schiller et al. (2018) – that are depleted in the noble gases and moderately volatile elements finalised Earth's accretion. In such a case, the process after disk dispersal would be a complex interplay between EUV-driven hydrodynamic and impact-induced atmospheric escape combined with mass accretion. Such a fast accretion scenario also fits with the Hf-W chronometry for $f_{Hf/W} \approx 15$, if full metal-silicate equilibrium is assumed (Rudge et al., 2010; Yu and Jacobsen, 2011; Kleine and Walker, 2017) (see also Sect. 3.1.1).

To conclude, the controversial hypothesis of Schiller et al. (2018) may give some answers as to why different Solar System bodies have varying compositions if they grew from the accretion of the same flow of pebbles (cf. Morbidelli, 2018), but it neither agrees with the isotopic nature of Earth's accreting material, as inferred by Dauphas (2017), (see also Sect. 3.2.1) nor are the implications clear for other isotopic systems and for the elemental abundances within the Earth. Future research will have to show how the $\mu^{48}Ca$ isotope data connects with the other isotopical, chemical, chronological and dynamical constraints of terrestrial planet formation.

### 3.2.3 Constraining early Earth's and Venus' accretion by reproducing their atmospheric $^{36}Ar/^{38}Ar$, $^{20}Ne/^{22}Ne$, $^{36}Ar/^{22}Ne$ isotope ratios

Besides the before mentioned dating methods, $^{36}Ar/^{38}Ar$, $^{20}Ne/^{22}Ne$ and $^{36}Ar/^{22}Ne$ noble gas isotope ratios in Earth's and Venus' atmospheres also provide important information on the origin and evolution of these planets (e.g. Pepin, 1991; 1997; Gillmann et al., 2009; Odert et al., 2018; Lammer at al., 2020a, this issue; Lammer et al., 2020b). Previous studies (e.g., Pepin 1991; 1997; Gillmann et al., 2009; Odert et al., 2018) that have tried to reproduce the present atmospheric isotope ratios of these noble gases during the past-applied arbitrary assumptions related to the EUV flux of the young Sun, simple atmospheric escape models and initial $H_2$-rich reservoirs related to primordial and steam atmospheres.

Marty (2012) analysed the Ar and Ne abundances in the present atmosphere of the Earth and found that the ratios of today's atmosphere and concentrations of water, carbon and nitrogen indicate a volatile content equivalent to ≈ 98% non-chondritic depleted/dry proto-Earth with only ≈ 2% contribution of carbonaceous chondritic material. As discussed above, Dauphas (2017) analysed isotopic signatures of lithophile, moderately siderophile and highly siderophile elements from Earth's mantle material and concluded that roughly 5% carbonaceous chondritic material was involved in early Earth's accretion. As discussed above, more recently, Schiller et al. (2018) inferred from the analysis of $^{48}$Ca/$^{44}$Ca isotope ratios in various early Solar System objects, that proto-Earth accreted roughly 40% carbonaceous chondrites (mainly CI chondrites) at the time when the disk dissipated. This value is roughly comparable to that obtained by Warren (2011), who uses the lever method to obtain ≈ 24 - 32 %. In the case that proto-Earth accreted higher amounts of carbonaceous chondritic material compared to the 2% inferred from today's data (Marty, 2012), thermal processing in a possible primordial atmosphere and atmospheric escape from accreting planetary embryos and the growing protoplanet could have been depleted in these volatile elements (Odert et al., 2018; Benedikt et al., 2019; Young et al., 2019; Lammer et al., 2020a, this issue; 2020b), altering the bulk and isotopic composition.

Recently, Lammer et al. (2020b) reproduced Earth's present atmospheric $^{36}$Ar/$^{38}$Ar, $^{20}$Ne/$^{22}$Ne and $^{36}$Ar/$^{22}$Ne noble gas isotope ratios within realistic young Sun and initial $H_2$-reservoir conditions by applying a sequence of sophisticated impact and upper atmosphere escape models.

accumulate around protoplanets with different masses during the accretion in the disk phase. They applied a hydrodynamic upper atmosphere model for the escape calculation of possible $H_2$-envelopes that remained during disk dispersal on various protoplanetary masses, including impact erosion, delivery, and composition/ratio modifications. The atmosphere evolution studies were carried out by including the young Sun's EUV flux constrained by an activity-rotation relation of young solar-like stars according to Tu et al. (2015).

Finally, Lammer et al. (2020b) could reproduce Earth's present $^{36}$Ar/$^{38}$Ar, $^{20}$Ne/$^{22}$Ne and $^{36}$Ar/$^{22}$Ne and Venus' $^{36}$Ar/$^{38}$Ar, $^{20}$Ne/$^{22}$Ne noble gas ratios within the uncertainties of the parameter space, when proto-Venus and proto-Earth grew to masses of up to 0.68 - 0.81$M_{Earth}$ and 0.53 – 0.58 $M_{Earth}$, respectively, at 4 Myr when the disk dissipated.
In early Earth's case, today's atmospheric $^{36}$Ar/$^{22}$Ne ratio of 18.8 (Marty and Allé, 1994; Marty, 2012) can be best reproduced by Lammer et al. (2020a; 2020b) if the post-nebula impactors (0.15 – 0.3 $M_{Earth}$) contain roughly 5% carbonaceous chondritic material (see Fig. 5; dashed lines) if one assumes an initial composition as suggested by Dauphas (2017), and ≥ 70% (see Fig. 5; solid lines) if they assume the composition as expected by Schiller et al. (2018).
The 95 % and ≤ 30 % of the accreted post-nebula material has to be depleted in the studied elements and hence originates from volatile-poor material.

These authors studied the boil-off phase of primordial $H_2$-dominated atmospheres that

Figure 5 shows reproduction attempts of Earth's present atmospheric Ar and Ne noble gas ratios when the proto-Earth was 0.55$M_{Earth}$ when the disk dissipated at 4 Myr. In such a case, a tiny primordial H$_2$-dominated atmosphere with a mass fraction of 0.0035 % $M_{Earth}$ remained after the boil-off phase during disk dispersal (Lammer et al., 2020a, this issue; 2020b).

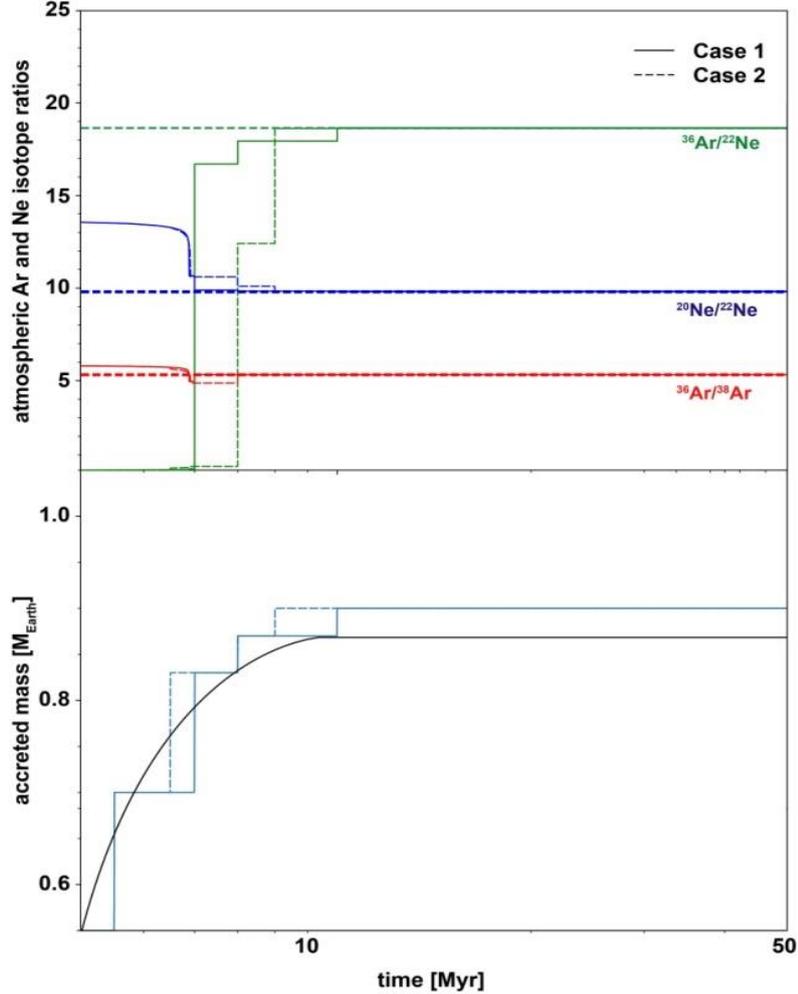

**Fig. 5:** Evolution scenario that reproduces present Earth's atmospheric $^{20}$Ne/$^{22}$Ne, $^{36}$Ar/$^{38}$Ar and $^{36}$Ar/$^{22}$Ne ratios and the corresponding early mass accretion of proto-Earth as modelled by Lammer et al. (2020a) with a proto-Earth-mass at 4 Myr of 0.55 $M_{Earth}$, a small captured H$_2$-dominated primordial atmosphere that is lost within < 7 Myr after the origin of the Solar System, and ≈ 3 Myr after disk dispersal due to impact erosion and EUV-driven hydrodynamic escape related to a young Sun that rotates slower than a moderate rotator. *Case 1*: Reproduction of Earth's atmospheric $^{20}$Ne/$^{22}$Ne, $^{36}$Ar/$^{38}$Ar and $^{36}$Ar/$^{22}$Ne ratios if one assumes the initial composition given in Dauphas (2017), but with a CC contribution of about 5 % after a tiny primordial atmosphere escaped (see also Sect. 3.2.1) with a contribution of ≈ 5% of carbonaceous chondrites. *Case 2*: Reproduction of Earth's atmospheric $^{20}$Ne/$^{22}$Ne, $^{36}$Ar/$^{38}$Ar and $^{36}$Ar/$^{22}$Ne ratios by assuming the initial composition suggested by Schiller et al. (2018) (see also Sect. 3.2.2) with a contribution of ≈ 40% of carbonaceous chondrites. The chondritic post-H$_2$-envelope impactors are depleted in 70% and 30% in Ne and Ar, respectively. The corresponding accretion scenario agrees well with that of Yu and Jacobsen (2011) that the main stage of Earth's accretion of ≈ 0.87 $M_{Earth}$ was completed rapidly within 10.7 ± 2.5 Myr, corresponding to a Hf-W determined Moon giant impact time $t_{MGI}$ at ≈ 53 Myr after Solar System formation (black solid line). (after Lammer et al., 2020a)

The growing proto-Earth lost this tiny primordial atmosphere during ≤ 3 Myr afterward by impact erosion and EUV-driven hydrodynamic escape of H atoms that drag and fractionate the Ar and Ne isotopes.

In Venus' case, the reproduction of its atmospheric $^{36}$Ar/$^{38}$Ar and $^{20}$Ne/$^{22}$Ne isotope ratios is more uncertain due to the large error bars of the available measurements of previous spacecraft. According to Lammer et al. (2020b) Venus' present atmospheric Ar and Ne isotope ratios can be reproduced if it captured a primordial atmosphere that corresponds to a mass for proto-Venus of ≈ 0.83 – 1 $M_{Venus}$ at the time when the disk dissipated. If this were the case then this would indicate that Venus accreted faster than Earth. As pointed out by Dandouras et al. (2020, this issue), future Venus missions that will precisely re-measure the atmospheric noble gas ratios are necessary to constrain the initial composition and evolution of Venus.

### 3.2.4 Constraining early Earth's accretion by $^3$He, and $^{20}$Ne/$^{22}$Ne, isotope data from the deep mantle

The question of whether the growing proto-Earth was surrounded by at least a low-mass primordial atmosphere has long been debated (e.g., Hayashi et al., 1979; Mizuno et al., 1980; Jacobsen and Harper, 1996; Harper and Jacobsen, 1996; Porcelli et al., 2001; Dauphas, 2003; Yokochi and Marty, 2004; Jacobsen et al., 2008; Raquin and Moreira, 2009; Hallis et al., 2015; Jaupart et al., 2017; Péron et al., 2018; Williams and Mukhopadhyay, 2019). That early Earth most likely accreted partially before and after the nebular disk dispersed was discussed by Harper and Jacobsen (1996), who studied the abundance of $^3$He and the before mentioned Ne isotopes in the mantle. In such a two-stage accretion scenario hypothesis, proto-Earth accreted enough mass during the disk lifetime to attract a $H_2$-He-dominated primordial atmosphere directly from the nebular gas. For explaining the evidence of the solar-like Ne and He noble gas abundances, as well as the presence of a large $^{129}$Xe excess in the deep Earth, Harper and Jacobsen (1996) and Jacobsen et al. (2008) suggest, that these elements were embedded in the solar-like $H_2$-He-dominated primordial atmosphere, dissolved in the magma ocean and transported into the deep Earth. Additionally, planetesimals and dust continuously added noble gas components into the growing protoplanet. Although, the presence of a large $^{129}$Xe excess, produced in the deep mantle due to a fast radioactive decay of $^{129}$I with a half-life of 15.7 Myr, is consistent with a very early atmosphere (Jacobsen and Harper, 1996; Jacobsen et al., 2008; Caracausi et al., 2016; Avice et al., 2017), the initially produced $^{129}$Xe is contaminated by Xe isotopes that were delivered during the late veneer. A detailed discussion on Xe isotopes and their perspectives on atmospheric evolution on Earth, Venus and Mars can be found in Avice and Marty (2020, this issue).

According to Harper and Jacobsen (1996), the inferred $^3$He abundance from deep-mantle plumes of Hawaiian volcanos (e.g., Poreda and Craig, 1992; Harper and Jacobsen, 1996; Timmerman et al., 2019) is more or less comparable to the abundance observed in primitive carbonaceous chondrites (e.g., Mazor et al., 1970; Kellogg and Wasserburg, 1990), but is much higher compared to all the other highly volatile elements in the deep Earth. This evidence is thought to be a primitive un-degassed reservoir that was preserved from the early accretion stages, when the planetary body was not large enough to support the energy of the impacts over the degassing threshold (Atreya et al., 1989; Harper and Jacobsen, 1996; Jacobsen et al., 2008).

On the other hand, one can expect that Earth's deep mantle should also be depleted in $^3$He in a similar way as the other volatiles, because of volatile depletion in proto-Earth's nebula source reservoir, metamorphic degassing in planetesimals, as well as degassing from large planetesimals and planetary embryos during accretion (Harper and Jacobsen, 1996; Jacobsen et al., 2008; Lichtenberg et al., 2016; 2018; Benedikt et al., 2020; Lammer et al., 2020a, this issue). Short-lived radioactive isotopes such as $^{26}$Al, and $^{60}$Fe also determine the thermal history and interior structure of the early planetary building blocks additionally to impacts and gravitational energy (Lichtenberg et al., 2016; 2018). The following thermo-mechanical evolution results in internal differentiation, which is related to a rapid volatile outgassing that was recently studied in Benedikt et al. (2020) and also discussed in Lammer et al. (2020a, this issue).

Therefore, the primordial $^3$He, which was accreted to the Earth through the delivery of carbonaceous chondrites, should have been efficiently lost, similar as the other volatile elements, compared to its initial abundance in the primitive meteorite value. By comparing the $^3$He abundance with other highly volatile elements that should have been accreted from the same building blocks, Harper and Jacobsen (1996) argued that the inferred $^3$He abundance in the deep Earth is at least two orders of magnitude higher than that expected from accretion of dust and planetesimals, which was probably most likely ≤ 1% of its initial value (Harper and Jacobsen, 1996; Jacobsen et al., 2008). Because of this paradox, these authors concluded that the $^3$He in Earth's deep mantle cannot be an accreted primitive component but is most likely a remnant of a magma ocean equilibration-related primitive $H_2$-He-dominated primordial atmosphere.

The preservation of such a primordial helium reservoir over the Earth's history has been questioned on the basis of geophysical evidence of slab subduction into the lower mantle and mantle convection models (e.g., van der Hilst et al., 1997; Maibom et al., 2003; Class and Goldstein, 2005; Parman, 2007). Recent studies, however, confirmed such a reservoir of primordial helium from studies of the hottest mantle plumes (Jackson and Becker, 2017) and from measured He-Sr-Pb isotope ratios in superdeep diamond fluid inclusions from the transition zone in a depth of 410 - 660 km which is unaffected by degassing and shallow crustal contamination (Timmerman et al., 2019). These authors conclude that their findings are indications that a less degassed, high-$^3$He/$^4$He deep mantle source infiltrates this transition zone, where it interacts with recycled material, creating the diverse compositions recorded in ocean island basalts. This is, furthermore, an indication of a dense high-$^3$He/$^4$He deep-mantle domain that could have remained isolated from the convecting mantle (e.g., Samuel and Farnetani, 2003), which explains and supports the preservation of very early (> 4500 Ma) geochemical anomalies in lavas (e.g., Harper and Jacobsen, 1996; Jacobsen et al. 2008; Mukhopadhay, 2012; Jackson et al., 2016).

Harper and Jacobsen (1996) suggested that the best explanation for the high abundance of $^3$He isotopes in the deep mantle is that 0.8 $M_{Earth}$ accreted within the presence of a solar nebular-composition atmosphere, which indicates a fast accretion of the early Earth. Such a scenario would be in agreement with the constrains obtained by the atmospheric Ar and Ne noble gas ratios (Lammer et al., 2020b), the Hf-W chronometer (Kleine and Walker, 2017, and references therein), and the Ca isotope systematics (Schiller et al., 2018), if proto Earth grew within the disk ($\approx$ 3.3 – 4.5 Myr; Bollard et al., 2017; Wang et al., 2017) to a mass of $\approx$ 0.53 - 0.58 $M_{Earth}$, and to $\approx$ 0.8 $M_{Earth}$ immediately after disk dispersal, whilst the growing body was surrounded by the escaping $H_2$-He-dominated primordial envelope (see Fig. 5).

Besides $^3$He, there is also an evidence of solar-like Ne isotopes in the deep mantle (e.g., Mizuno et al., 1980; Honda et al., 1991; Porcelli et al., 2001; Yokochi and Marty, 2004; Jacobsen et al., 2008; Raquin and Moreira, 2009; Jaupart et al., 2017; Péron et al., 2018). Williams and Mukopadhayay (2019) measured a $^{20}$Ne/$^{22}$Ne isotope ratio from Earth's deep interior of 13.01 – 13.45, which is higher than a ratio of 12.52-12.75 that could also be produced through implantation of the solar wind onto accreted meteoritic material (Moreira and Kurz, 2013; Moreira and Charnoz, 2016). This range is well in accordance with the $^{20}$Ne/$^{22}$Ne isotope ratio of $\approx$13.34 of the solar nebula that was derived by Heber et al. (2012) from Genesis data. In agreement with the hypothesis of Harper and Jacobsen (1996) and Jacobsen et al. (2008), Williams and Mukhopadhayay (2019) suggest that some Ne isotopes that were embedded in a primordial atmosphere were incorporated with the solar ratio into proto-Earth's interior via a magma ocean. Jacobsen et al. (2008) also argued that solar wind related implanted components would most likely be thoroughly outgassed and lost from early differentiated planetesimals due to the before mentioned $^{26}$Al heating (Lichtenberg et al., 2016; 2018), which comports with the results of recent studies such as Young et al. (2019), Sossi et al. (2019), Benedikt et al. (2020) and Lammer et al. (2020a, this issue)

Slower accretion scenarios with proto-Earth accreting $< 0.53$ $M_{Earth}$ during the disk lifetime of 3.3 – 4.5 Myr (Bollard et al., 2017; Wang et al., 2017) would need planetary embryos that were extremely depleted in volatile and moderately volatile rock-forming elements (i.e. K, Na, Si, Mg, etc.) after the disk evaporated and other explanations than a primordial atmospheric origin for the high $^3$He abundance in the deep mantle would have to be found.

If proto-Earth, on the other hand, accreted too much mass during the disk lifetime (i.e., a mass of $> 0.7$ $M_{Earth}$), an H$_2$-dominated primordial atmosphere with a mass of $\geq 0.0042$ $M_{Earth}$ would remain after disk dispersal, which could not be lost by EUV-driven hydrodynamic escape during a billion years, so that Earth would have most likely become a sub-Neptune-like body.

### 3.2.5 Containing early Earth's accretion from D/H

If proto-Earth grew partly within the gas disk then one can argue that there should be visible evidence in the D/H ratio of Earth's ocean water (e.g. Meech and Raymond, 2019; and references therein; Raymond and Morbidelli, 2020), since the D/H ratio of nebular gas of $(21 \pm 5) \times 10^{-6}$ is roughly 7 times lower (Geiss and Gloeckler, 1998; Robert et al., 1999) than Earth's seawater D/H ratio of $(150 \pm 10) \times 10^{-6}$ (e.g., Robert et al. 1999), which fits well within the measured D/H ratio in carbonaceous chondrites (e.g., Marty 2012). According to Ikoma and Genda (2006) and Ikoma et al. (2018), the atmospheric H$_2$ within the primordial H$_2$-dominated atmosphere can be oxidized by gas-rock interactions between the atmosphere and a magmatic surface with temperatures $\geq 1500$ K, thereby producing H$_2$O. The amount of nebular-related water depends both on the mass of the surrounding primordial atmosphere and on the available ion oxides and fayalite (Fe$_2$SiO4) in the magma ocean. The ratios of produced water to the primordial atmospheric hydrogen $M_{H2O}/M_{H2}$ are $\approx 0.49$, $\approx 24.02$, and $\approx 0.88$ for the quarz-iron-fayalite-oxygen (2Fe + SiO$_2$ + O$_2$ $\leftrightarrow$ Fe$_2$SiO$_4$), the wüstite-magnetite (6.696Fe$_{0.974}$O + O$_2$ $\leftrightarrow$ 2.174Fe$_3$O$_4$), and the ion-wüstite (1.894Fe + O$_2$ $\leftrightarrow$ 2Fe$_{0.947}$O) buffers (Robie et al., 1978; Ikoma and Genda, 2006). Ikoma and Genda (2006) estimated that under such conditions a protoplanet that captured an H$_2$-envelope from the disk gas with a mass of $M_{H2} \approx 10^{22}$ kg (0.00167 $M_{Earth}$) would produce an H$_2$O amount comparable to the mass of the Earth's current ocean of $M_{H2O} \approx 1.4 \times 10^{21}$g if such oxygen buffers were available.

By applying the same assumption as Ikoma and Genda (2006) for an accretion scenario that can reproduce the present day atmospheric Ar and Ne isotope ratios based on the simulations of Lammer et al. (2020a), a proto-Earth with a mass of 0.55 $M_{Earth}$ (see also Fig. 5) with a captured primordial H-dominated atmosphere of roughly $M_{H2} \approx 3.5 \times 10^{-5} M_{Earth}$ (or $2.15 \times 10^{20}$ kg) produces a water reservoir of $M_{H2O}$ of $\approx 3.0 \times 10^{19}$ g from the $H_2$-envelope corresponding to only $\approx 2\%$ of the current value of Earth's seawater (see Fig. 6). Such a low value lies within the error bars of the D/H seawater-carbonaceous chondrite "match" which is $\approx 10 - 20\%$ (Pahlevan et al., 2019; and references therein).

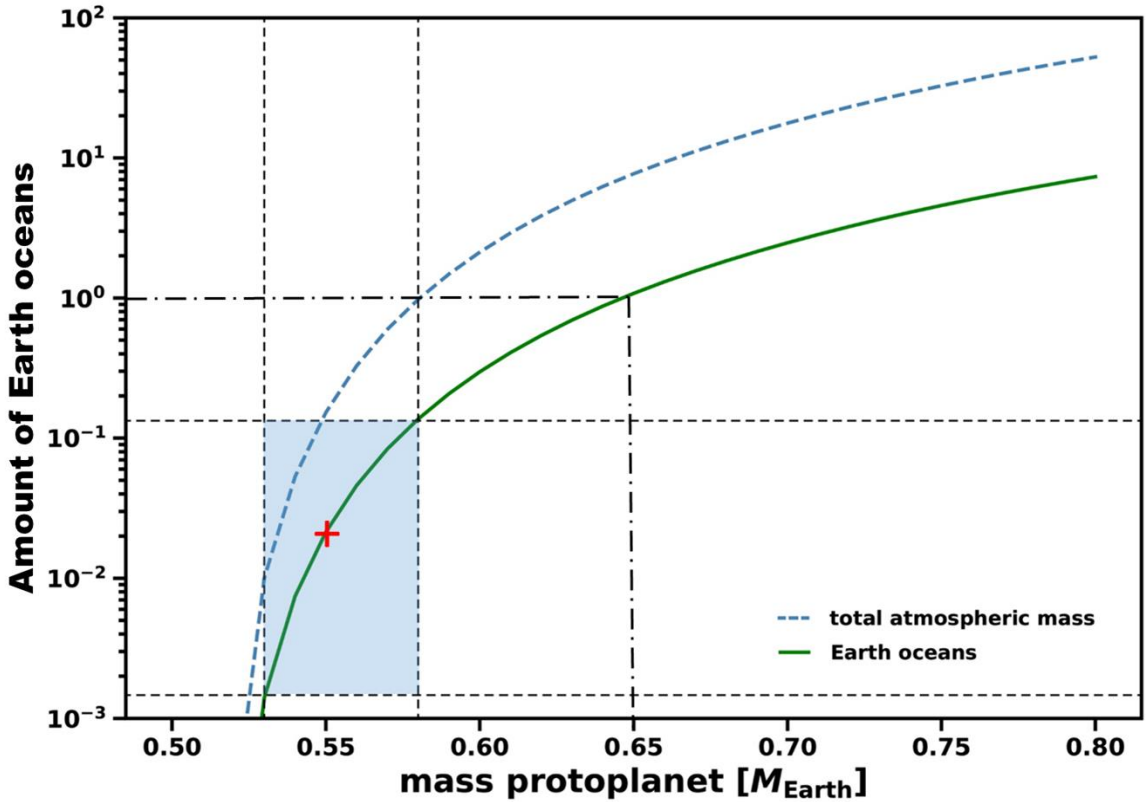

**Fig. 6:** The amount of Earth oceans (green solid line) that are produced by the captured $H_2$-dominated primordial atmospheres (blue dashed line) as modelled in Lammer et al. (2020b) due to gas-rock interactions between the atmosphere and a magma ocean according to Ikoma and Genda (2006). The red "+" corresponds to the accretion scenario that can reproduce Earth's atmospheric Ar and Ne isotope ratios, as shown in Fig. 5. In this scenario, a negligible amount of 2% of an Earth ocean with solar nebula origin would be produced. The blue area corresponds to the proto-Earth masses at the disk dissipation time for which the nebula-based D/H-ratio would only modify the chondritic value of Earth's seawater within its error bars. The dashed-dotted lines mark the production of an Earth ocean from a primordial in case of proto-Earth accreting to 0.65 $M_{Earth}$ before the gas disk dissipated.

If we estimate the hypothetical $H_2O$ production with similar assumptions as before but for a proto-Earth with a mass of 0.75 $M_{Earth}$ at the disk dispersal time of 4 Myr, an $H_2$-envelope of 0.0077 $M_{Earth}$ ($4.59 \times 10^{22}$ kg) would be captured (Lammer et al., 2020b). In such a case, roughly 4.5 times of Earth's present seawater value, or some 4.5 Earth-oceans, could have been produced due to the nebula gas-magma ocean interaction processes. One can see from Fig. 6, that proto-Earth masses > 0.6 $M_{Earth}$ should have left behind D/H ratios that would differ largely from the measured carbonaceous chondritic value.

Another indication that at least a small fraction of Earth's H2O may be derived from a nebular origin comes from measurements of the D/H ratios in glassy melt inclusions in two basaltic rock samples that originated in a deep mantle reservoir by Hallis et al. (2015). These authors found D/H ratios in their samples that are closer to the nebular than to the carbonaceous chondritic ratio.

According to the reproduction attempts of Venus' atmospheric Ar, Ne and bulk K/U ratios (Lammer et al., 2020a, this issue; 2020b), more nebula-based water could have been produced compared to early Earth's expected evolution as shown in the blue area of Fig. 6 due to proto-Venus possibly being slightly more massive. This would indicate that Venus' initial D/H ratio was probably closer to the nebula ratio than to carbonaceous chondrites. Since the planet's initial water inventory and its D/H ratio was modified during the planets evolution by thermal escape and various nonthermal atmospheric loss processes as well as water vapor outgassing episodes (see Lammer et al., 2020a, this issue) it is difficult to draw definite conclusions. New future in-situ measurements of Venus atmospheric D/H ratio in combination with heavier noble gas isotopes as discussed in Dandouras et al. (2020; this issue) are essential for constraining and understanding the evolution of early Venus' water inventory and atmosphere.

One should also note that the production of nebula-based water depend on the nebula properties such as the dust grain depletion factor, etc. (Lammer et al., 2014; Kimura and Ikoma, 2020), which can be different on other star systems. Recently, Kimura and Ikoma (2020) showed that primordial atmospheres could be highly enriched in water vapor, if one does not assume solar abundances for the gas disk properties.

### 3.6 Protoplanetary masses of Venus, Earth and Mars at disk dispersal time constrained from different isotope reproduction studies

If we summarize the different discussed isotope-systematics that are related to D/H, primordial $^{3}$He, $^{36}$Ar/$^{38}$Ar, $^{20}$Ne/$^{22}$Ne, $^{36}$Ar/$^{22}$Ne, $^{48}$Ca/$^{44}$Ca abundances and ratios, and chronometric methods based on $^{182}$Hf-$^{182}$W and U-P isotopes one obtains the accretion scenarios for early Earth illustrated in Fig. 7.

The min. and max. disk lifetime according to Bollard et al. (2017) and Wang et al. (2017) is represented by the two vertical dashed-lines. The lower horizontal dashed-line corresponds to the minimum proto-Earth-mass, which is necessary for a protoplanet at 1 AU to build-up and keep an $H_2$-He-dominated primordial atmosphere (Lammer et al., 2020b) from the solar nebula for some while. The blueish shaded area above the mass fraction of proto-Earth corresponds to masses for which primordial atmospheres remain too long around the growing planetary nucleus for a distance of 1 AU (Sasaki and Nakazawa, 1990; Lammer et al. 2020a, this issue; 2020b). As discussed in Sec. 3.2.4, there is an excess of $^3$He in the deep mantle, which is two orders of magnitude higher than the $^3$He amount that was delivered by planetesimals and dust. Because it is expected that this $^3$He excess originated most likely due to ingassing (equilibration with a global magma ocean) from a primordial $H_2$-He-dominated atmosphere (e.g., Harper and Jacobsen, 1996; Jacobsen et al., 2008; Jackson et al., 2017; Timmerman et al., 2019), one can use this primordial $^3$He evidence for constraining the minimum proto-Earth-mass at the disk dispersal time, in agreement with the reproduced atmospheric $^{36}Ar/^{38}Ar$, $^{20}Ne/^{22}Ne$, $^{36}Ar/^{22}Ne$, noble gas ratios (see Sect; 3.2.3, Lammer et al., 2020a; 2020b, this issue) of 0.53 $M_{Earth}$.

The upper mass of proto-Earth at the disk dispersal time can be constrained with the D/H ratio of Earth's seawater in agreement with the reproduction of the atmospheric Ar and Ne isotope ratios (see Sect. 3.2.3; Lammer et al., 2020a; this issue; 2020b) of $0.5 < M \leq 0.6$ $M_{Earth}$.

As discussed in Sect. 3.2.5, higher proto-Earth-masses would have resulted in too massive $H_2$-He-dominated primordial envelopes that would have produced a too high amount of ocean water through the reaction between the primordial atmospheric hydrogen and oxides in the magma ocean (Ikoma and Genda, 2006). Such a mixture of water from nebular and carbonaceous chondritic origin would have resulted in different D/H ratios than measured in Earth's seawater, which is close to the D/H value measured in carbonaceous chondrites (e.g., Marty 2012). The dark grey box that is confined between proto-Earth masses of $\approx$ 0.53 - 0.6 $M_{Earth}$, and the min and max disk dispersal time of 3.3 and 4.5 Myr, therefore, corresponds to the range to which the accreting proto-Earth should have evolved to produce the preserved isotope records. The dotted-vertical line marks the time when the primordial atmosphere was lost (Lammer et al., 2020a).

In case proto-Earth accreted very fast (dotted brownish line), almost to its whole mass within 6 - 7 Myr. Schiller et al. (2018) inferred it from Ca-isotope systematics (see Sect. 3.2.2), then the disk lifetime should have been close to its minimum value. At this time, the luminosity of the young Sun was yet higher so that a slightly denser primordial atmosphere could have been lost from a more massive proto-Earth of 0.6 $M_{Earth}$ at the time of disk dispersal. A continuous fast growth until about 7 Myr after the origin of the Solar System combined with the escaping primordial atmosphere can reproduce the present atmospheric Ar and Ne isotope ratios including $^{36}Ar/^{22}Ne$ (see Lammer et al., 2020a for details).

The solid and dashed brownish lines correspond to the dual or two stage accretion scenarios as proposed by Yu and Jacobsen (2011) and Dauphas (2003). Both scenarios can also produce the studied isotope record. Most of the recent studies that used the Hf-W-chronometry method for dating the core formation expect a complete equilibrium or nearly complete equilibrium of metal and silicates favouring Hf/W values betweeb 10 – 20, thereby indicating a fast accretion of $\geq 0.8\ M_{Earth}$ before $\leq 30$ Myr (Rudge et al. 2010; Kleine et al., 2002; Yin and Jacobsen, 2006; Yu and Jacobsen, 2011; Kleine and Walker, 2017; and references therein). There are difficulties to homogenise isotopic data related to the composition of Earth's accreted material between the measurements of lithophile, and moderate and highly siderophile elements (Dauphas, 2017) and $\mu^{48}Ca$ (Schiller et al., 2018).

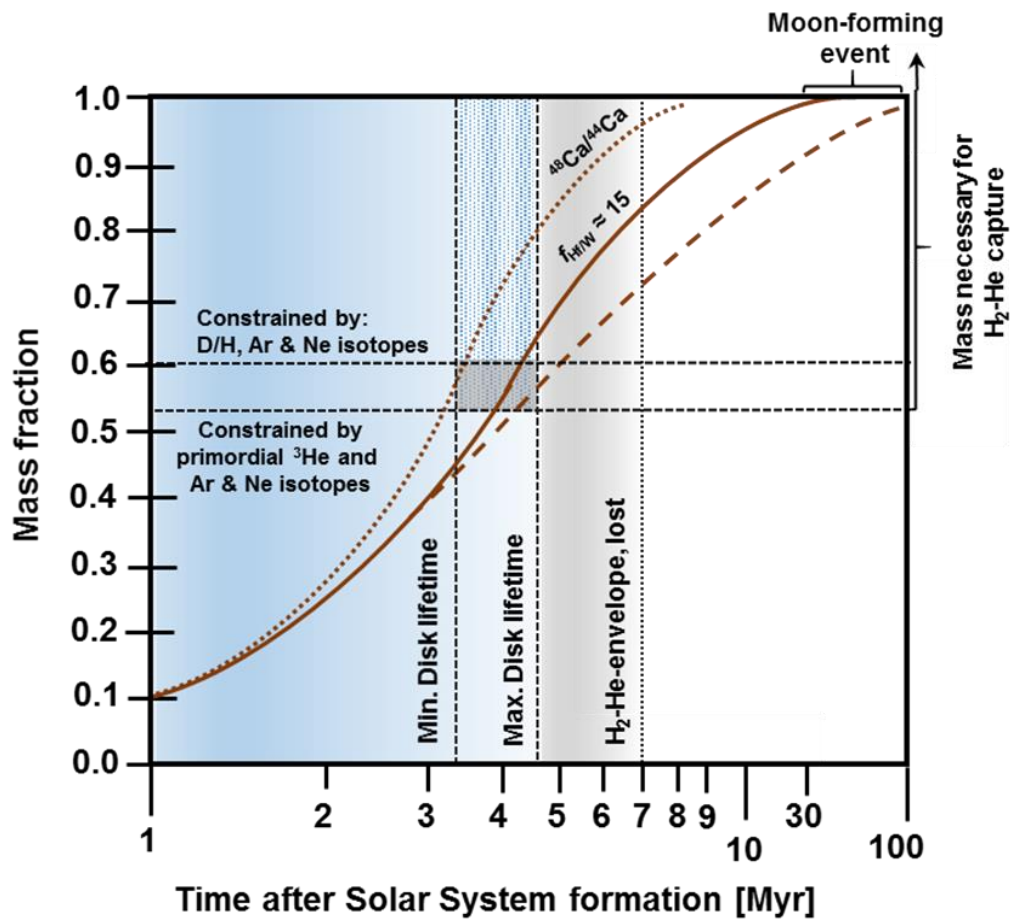

Fig. 7: Illustration of the most likely proto-Earth accretion scenarios as constrained by different isotope-systematics (D-H, atmospheric Ar & Ne, primordial $^3$He abundance in the deep mantle) and isotopic chronometers (Hf-W, U-P) in dependence on the disk lifetime and activity of the young Sun. Proto-Earth's mass fraction during disk dispersal should have been $\approx 0.5 – 0.6\ M_{Earth}$ (dark grey area).

From this discussion, one can see that the initial setup of planets that may later evolve to an Earth-like habitat is a complex interplay between the lifetime of the gas-disk, the accretion rate of the terrestrial planet and the EUV evolution path of its young host star. If the Solar System disk lifetime had been about 8 Myr instead of ≈ 4 Myr (Bollard et al., 2017; Wang et al. 2017), Earth would have accreted too much primordial atmospheric mass within the protoplanetary nebula and would have ended as a small sub-Neptune or an ocean planet surrounded by a remnant of its primordial atmosphere that could not escape during its lifetime.

Due to insufficient noble gas isotope data, at present one cannot conclude much about the main composition of proto-Venus. If, however, proto-Venus accumulated a primordial atmosphere, it should have finished its accretion slightly faster than proto-Earth and did not grow further due to a lack of planetary embryos/accreting material at 0.7 AU, while early Earth at 1 AU could have still accreted ≤ 0.4 $M_{Earth}$ after the disk dispersal (Lammer et al., 2020a; 2020b, this issue).

In the case of early Mars, according to Hf-W chronometry (Dauphas and Pourmand, 2011; Kleine and Walker, 2017), this small planet also accreted most of its mass (> 0.63 – 1.0 $M_{Mars}$) during the disk phase, but proto-Mars was too small to capture a tiny primordial atmosphere that remained around its rocky nucleus after the dissipation of the disk (Lammer et al., 2018).

Out of the formation hypotheses and model results that are compared in Table 1 of Section 2, currently the most likely ones to have formed terrestrial planets are pebble accretion (Johansen and Lambrechts, 2017) and some Grand Tack models (Brasser et al., 2016); they come the closest to the accretion of planetary embryos with masses of ≈ 0.5$M_{Earth}$ after a few Myr. Planetesimals grow into large planetary embryos, even giant cores, or gas giants by accreting planetesimals or pebbles, or most likely a combination of both (Lambrechts et al., 2019). One should note that model simulations of planetesimal accretion struggle to grow giant massive planetary cores within the disk lifetime of 3-5 Myr (e.g., Levison et al., 2015a; 2015b). Such fast accretion scenarios may be solved by pebble accretion, but key aspects of pebble accretion remain to be better understood (Lambrechts and Johansen, 2012; Johansen et al., 2015; 2017; Birnstiel et al., 2016; Bitsch et al., 2019).

According to studies of Johansen et al. (2015), recent simulations indicate that the main growth of planetesimals results from the gas drag-assisted accretion of chondrules. Thus, chondrules represent the building blocks of planetesimals, planetary embryos, protoplanets and finally, terrestrial planets (e.g., Lambrechts and Johansen, 2012; Johansen et al., 2015; Birnstiel et al., 2016; Johansen and Lambrechts, 2017; Alexander, 2019a; 2019b). It is expected that these scenarios evolve from a two-stage process, where the first generation of accreting bodies form very fast by streaming instabilities followed by continuous growth that is dominated by the gas-drag in combination with the accretion of mm-sized particles for planetesimals with radii larger than about 200 km (Johansen et al., 2015). The streaming instability most likely represents a promising mechanism that can explain how dust particles grow to 100 km large planetesimals. The streaming instability, however, requires specific conditions to operate and this implies that planetesimals may form in preferential locations, just beyond the snow line (Izidoro and Raymond, 2017). However, at present knowledge, pebble accretion and streaming instabilities most likely solve the first stages in planet accretion (Fig. 1: I and II), as it results in the fast formation of Mars-sized embryos and even more massive protoplanets $\geq 0.5$ $M_{Earth}$ over typical disk lifetimes of 3 - 5 Myr. The later stage of accretion (Fig. 1: III, IV) of the terrestrial planet structure of the inner Solar System can then be quite well explained by the Grand-Tack model and to some extent by the annulus and depleted disk models as well. A clear future step in terrestrial planet formation will be an N-body simulation that also includes the reproduction of isotopic and geochemical constraints.

## 4 Conclusions

We reviewed the classical scenarios of terrestrial planet formation, including the Grand-Tack model and more recently proposed alternatives. The terrestrial planet formation models were compared with accretion time scales obtained from isotope data of planetary building blocks, the atmospheric Ar and Ne isotope data of Earth and Venus, Earth's D/H isotope ratio in the sea water and the evidence of $^3$H abundance in the deep mantle, as well as $^{182}$Hf-$^{182}$W chronometry. The analysis of these data indicate that the early terrestrial planets in the Solar System could have accreted most of their mass ($\geq 0.8$ $M_{Earth}$) within the first 10-30 Myr, including a Moon-forming event related to Earth $\geq 50$ Myr after the Solar System originated. From the combined evidence of the atmospheric $^{36}$Ar/$^{38}$Ar, $^{20}$Ne/$^{22}$Ne, $^{36}$Ar/$^{22}$Ne isotope ratios, the D/H ratio in Earth's seawater and the estimated $^3$He abundance that likely originated due to ingassing from a primordial atmosphere, one finds that proto-Earth accreted a mass of 0.5 – 0.6 $M_{Earth}$ at $\approx 4$ Myr, the time when the disk dissipated. The planetesimals grow into planetary embryos and or protoplanets by accreting pebbles, planetesimals, or a combination of both. Pebble accretion most likely solves time scale conflicts of fast accretion, although some aspects of pebble accretion remain to be better understood. For instance, more observations of the radial dependence of pebble sizes in disks, coagulation models that include volatiles, and experiments on the sticking properties of aggregates of a range of compositions are necessary. The late stage accretion of terrestrial planets can than explain the structure of the inner Solar System. Here, the Grand Tack model, the primordial low-mass asteroid belt, and the primordial empty asteroid belt models yield quite accurate results. More studies are needed to homogenise pebble accretion and later stage accretion models with data that are provided by chemical, chronological, isotopical and dynamical constraints. Furthermore, precise measurements of the isotopic composition of Venus are necessary for testing the planet's evolution, including the isotopic composition of its building blocks.

**Acknowledgements** R. B. acknowledges financial assistance from the Japan Society for the Promotion of Science (JSPS) Shingakujutsu Kobo (JP19H05071). A.J. acknowledges funding from the European Research Foundation (ERC Consolidator Grant 724687-PLANETESYS), the Knut and Alice Wallenberg Foundation (Wallenberg Academy Fellow Grant 2017.0287) and the Swedish Research Council (Project Grant 2018-04867). We thank Nathan Kaib and Matt Clement for information regarding their 'Early instability' model. H. L. and M. S. acknowledge support from the FWF NFN subproject S11607-N16. M. L. acknowledge support of the Austrian FWF projects P27256-N27 and P30949-N36. The authors also thank the International Space Science Institute (ISSI) in Bern, Switzerland for the support. Finally, we thank the referee S. B. Jacobsen and an anonymous referee for their very helpful comments and suggestions that helped to improve this review article.

**References:**

Agnor, C.B., Canup, R.M., Levison, H.F., 1999. On the Character and Consequences of Large Impacts in the Late Stage of Terrestrial Planet Formation. Icarus, 142:219-237.

Ahrens, T. J., O'Keefe, J. D., Lange, M. A., 1989. Formation of atmospheres during accretion of planets. In: Origin and evolution of planetary atmospheres. (eds., S. K. Atreya, J. B. Pollack, M. S. Matthews), University of Arizona Press, Tucson, pp. 35-77.

Alexander C. M. O., 2019a. Quantitative models for the elemental and isotopic fractionations in chondrites: The carbonaceous chondrites. Geochim. Cosmochim. Acta, 254:277–309.

Alexander C. M. O., 2019b, Quantitative models for the elemental and isotopic fractionations in the chondrites: The noncarbonaceous chondrites. Geochim. Cosmochim. Acta, 254:246–276.

Allègre, C. J., Manhès, G., Göpel, C., 2008. The major differentiation of the Earth at ~4.45 Ga. Earth Planet. Sci. Lett., 267:386-398.

Albarède, F., 2009. Volatile accretion history of the terrestrial planets and dynamic implications. Nature, 461:1227-1233.

Asphaug, E., 2014. Impact origin of the Moon. Annu. Rev. Earth Planet. Sci. 42:551-578.

Avice, G., Marty, B., Burges, R., 2017. The origin and degassing history of the Earth's atmosphere revealed by Archean xenon. Nature Communications, 8:15455.

Avice, G., Marty, B., 2020. Perspectives on atmospheric evolution from noble gas and nitrogen isotopes on Earth, Mars & Venus. Space Sci. Rev., this issue.

Barboni, M., Boehnke, P., Keller, B., Kohl, I. E., Schoene B., Young, E D., McKeegan, K. D., 2017. Early formation of the Moon 4.51 billion years ago. Science Adv., 3:e1602365.

Benedikt, M. R., Scherf, M., Lammer, H., Marcq, E., Odert, P., Leitzinger, M., Erkaev, N.V. 2020. Escape of rock-forming volatile elements and noble gases from planetary embryos. Icarus, 347:113772.

Bennett, V. C., Brandon, A. D., Nutman, A. P., 2007. Coupled 142Nd–143Nd isotopic evidence for Hadean mantle dynamics. Science, 318:1907–1910.

Bermingham, K., Füri, E., Lodders, K., Marty, B., 2020. Chemical and isotopic evolution of the early Solar System. Space Sci. Rev., this issue.

Birnstiel, T., Fang, M., Johansen, A. 2016. Dust evolution and the formation of planetesimals. Space Sci. Rev., 205:41- 45.

Bitsch, B., Izidoro, A., Johansen, A., Raymond, S. N., Morbidelli, A., Lambrechts, M., Jacobson, S., 2019. Formation of planetary systems by pebble accretion and migration: growth f gas giants. Astron. Astrophys., 623:A88.

Bollard, J., Connelly, J. N., Whitehouse, M. J., Pringle, E. A., Bonal, L., Jørgensen, J. K., Nordlund, Å., Moynier, F., Bizzarro M., 2017. Early formation of planetary building blocks inferred from Pb isotopic ages of chondrules. Science Adv., 3:1-9.

Bouhifd, M. A., Jephcoat, A. P., 2011. Convergence of Ni and Co metal-silicate partition co-efficients in the deep magma-ocean and coupled silicon-oxygen solubility in iron melts at high pressures. Earth Planet. Sci. Lett., 307:341-348.

Boyet, M., Blichert-Toft, J., Rosing, M., Storey, M., Telouk, P., Albarede, F., 2003. $^{142}$Nd evidence for early Earth differentiation. Earth Planet. Sci. Lett. 214:427–442.

Boyet, M., Calson, R. W., 2005. 142$^{Nd}$ evidence for early (>4.53 Ga) global differentiation of the silicate Earth. Science, 309:576-581.

Brasser, R., Morbidelli, A., Gomes, R., Tsiganis, K., Levison, H.-F., 2009. Constructing the secular architecture of the solar system II: the terrestrial planets. Astronomy and Astrophysics, 507:1053-1065.

Brasser, R., 2013. The formation of Mars: Building blocks and accretion time scale. Space Sci. Rev., 174, 11-25.

Brasser, R., Lee, M.H., 2015. Tilting Saturn without tilting Jupiter: Constraints on giant planet migration. Astron. J., 150, ID 157.

Brasser, R., Matsumura, S., Ida, S., Mojzsis, S.J., Werner, S.C., 2016. Analysis of terrestrial planet formation by the grand tack model: system architecture and tack location. The Astrophys. J., 821, 75-93.

Brasser, R., Mojzsis, S.J., Matsumura, S., Ida, S., 2017. The cool and distant formation of Mars. Earth and Planet. Sci. Lett., 468: 85-93.

Brasser, R., Dauphas, N., Mojzsis, S.J., 2018. Jupiter's influence on the building blocks of Mars and Earth. Geophys. Res. Lett., 45: 5908-5917.

Brasser, R., Matsumura, S., Ida, S., Moizsis, S. J., Werner, S. C., 2019. Analysis of terrestrial planet formation by the Grand Tack model: System architecture and tack location. Astrophys. J., 821:75, 18pp.

Caracausi, A., Avice, G., Burnard, P. G., Füri, E., Marty, B., 2016. Chondritic xenon in the Earth's mantle. Nature, 533:82-85.

Carlson, R.W., 2017. Earth's building blocks. Nature, 541: 468-469.

Carlson, R.W., Brasser, R., Yin, Q.-Z., Fischer-Gödde, M., Qin, L, 2018. Feedstocks of the Terrestrial Planets. Space Sci. Rev., 214: article id. 121, 32 pp

Caro, G., Bourdon, B., Birck, J.-L., Moorbath, S., 2003. $^{146}$Sm–$^{142}$Nd from Isua metamorphosed sediments for early differentiation of the Earth's mantle. Nature, 423:428–431.

Chambers, J.E., 2001. Making more terrestrial planets. Icarus, 152:205-224.

Chambers, J.E., Wetherill, G.W., 1998. Making the terrestrial planets: N-body integrations of planetary embryos in three dimensions. Icarus, 136:304-327.

Class, C., Goldstein, S.L, 2005. Evolution of helium isotopes in the Earth's mantle. Nature, 436:1107–1112.

Clayton, D. D., 1970 Nucleosysnthesis in stars. Proceed. Of the 11$^{th}$ Int. conf. on cosmic rays, 5:21-39.

Clement, M.S., Kaib, N.A., Raymond, S.N., Chambers, J.E., Walsh, K.J., 2019. The early instability scenario: Terrestrial planet formation during the giant planet instability, and the effect of collisional fragmentation. Icarus, 321:778-790.

Connelly, J. N., Schiller, M., Bizzarro, M., 2019. Pb isotope evidence for rapid accretion and differentiation of planetary embryos. Earth Plant. Sci. Lett., 525:115722.

Cresswell, P., Nelson, R. P., 2008. Three-dimensional simulations of multiple protoplanets embedded in a protostellar disc. Astron. Astophys., 482:677-690.

Dandouras, I., Blanc, M., Fossati, L., Gerasimov, M., Guenther, E. W., Kislyakova, K. G., Lammer, H., Lin, Y., Marty, B., Rugheimer, S., Sotin, C., Tachibana, S., Wurz, P.,

Yamauchi, M., 2020. Future missions related to isotope and element measurements. Space Sci. Rev., this issue.

Dauphas, N., 2003. The dual origin of the terrestrial atmosphere. Icarus, 165, 326-339.

Dauphas, N., Pourmand, A., 2011. Hf-W-Th evidence for rapid growth of Mars and its status as a planetary embryo. Nature, 473:489–92.

Dauphas, N., 2017. The isotopic nature of the Earth's accreting material through time. Nature, 541:521-524.

D'Angelo, G., Marzari, F., 2012. Outward Migration of Jupiter and Saturn in Evolved Gaseous Disks. Astrophys. J. 757:50.

DeMeo, F.E., Carry, B., 2014. Solar System evolution from compositional mapping of the asteroid belt. Nature, 505:629-634.

Drake, M. J., Righter, K., Determining the composition of the Earth. Nature, 416:39-44.

Drazkowska, J., Alibert, Y., Moore, B., 2016. Close-in planetesimal formation by pile-up of drifting pebbles. Astron. Astrophys., 594, A105.

Fischer-Gödde, M., Kleine, T., 2017. Ruthenium isotopic evidence for an inner Solar System origin of the late veneer. Nature, 541:525-527.

Fossati, L., Erkaev, N. V., Lammer, H., Cubillos, P. E., Odert, P., Juvan, I., Kislyakova, K. G., Lendl, M., Kubyshkina, D., Bauer S. J., 2017. Aeronomical constraints to the minimum mass and maximum radius of hot low-mass planets. Astron. Astrophys., 598:90-99.

Geiss J., Gloeckler G., 1998. Abundances of Deuterium and Helium-3 in the Protosolar Cloud. Space Sci. Rev., 84:239-250.

Gillmann, C., Chassefière, E., Lognonné, P., 2009. A consistent picture of early hydrodynamic escape of Venus atmosphere explaining present Ne and Ar isotopic ratios and low oxygen atmospheric content. Earth Plant. Sci. Lett., 286:503-513.

Gomes, R., Levison, H.F., Tsiganis, K., Morbidelli, A., 2005. Origin of the cataclysmic Late Heavy Bombardment period of the terrestrial planets. Nature, 435:466-469.

Halliday, A. N., Lee, D. C., 1999. Tungsten isotopes and the early development of the Earth and Moon, Geochim. Cosmochim. Acta, 63:4157–4179.

Halliday, A. N., 2000. Hf-W chronometry and inner solar system accretion rates. Space Sci. Rev., 92:355-370, 2000.

Halliday, A. N., 2004. Mixing, volatile loss and compositional chance during impact-driven accretion of the Earth. Nature, 427:505-509.

Halliday, B. J. Wood, B. J., 2007. 9.02 The composition and major reservoirs of the Earth around the time of the Moon-forming giant impact in: Treatise on Geophysics: Evolution of the Earth. (Ed. D. Stevenson), 9:13-50.

Halliday, A. N., 2008. A young Moon-forming impact at 70-110 million years accompanied by late-stage mixing, core formation and degassing of the Earth. Phil. Trans. R. Soc. A 366:4163-4181.

Halliday, A. N., Wood, B. J., 2009. How did Earth accrete? Science, 325:44-45.

Hallis L. J., Huss G. R., Nagashima K., Taylor G. J., Halldórsson S. A., Hilton D. R., Mottl M. J., and Meech K. J., 2015. Evidence for primordial water in Earth's deep mantle, Science, 350:795–797.

Hansen, B.M.S., 2009. Formation of the Terrestrial Planets from a Narrow Annulus. Astrophys. J., 703:1131-1140.

Harper, C. L., Jr., Jacobsen S. B., 1996. Noble gases and Earth's accretion. Science 273:1814-1818.

Hartmann, L., Calvet, N., Gullbring, E., D'Alessio, P., 1998. Accretion and the evolution of T Tauri disks. The Astrophysical Journal, 495:385-400.

Hayashi, C., Nakazawa, K., Mizuno, H., 1979. Earth's melting due to the blanketing effect of the primordial dense atmosphere, Earth Planet. Sci. Lett., 43, 22-28.

Helled, R.; Bodenheimer, P.; Podolak, M.; Boley, A.; Meru, F.; Nayakshin, S.; Fortney, J. J.; Mayer, L.; Alibert, Y.; Boss, A. P., 2013. Giant Planet Formation, Evolution, and Internal Structure. In: Protostars and Planets VI, Henrik Beuther, Ralf S. Klessen, Cornelis P. Dullemond, and Thomas Henning (eds.), University of Arizona Press, Tucson, 914 pp., 643-665.

Honda, M., McDougall, I., Patterson, D. B., Doulgeris, A., Clague, D. A., 1991. Possible solar noble-gas component in Hawaiian basalts. Nature, 349:149-151.

Ida, S., Makino, J., 1993. Scattering of planetesimals by a protoplanet - slowing down the runaway growth. Icarus, 106:210-227.

Ikoma, M., Genda, H., 2006. Constraints on the Mass of a Habitable Planet with Water of Nebular Origin. Astrophys J., 648:696-706.

Ikoma, M., Elkins-Tanton, L., Hamano, K., Suckale, J., 2018. Water partitioning in planetary embryos and protoplanets with magma oceans. Space Sci. Rev., 214:76

Izidoro, A., Haghighipour, N., Winter, O. C., Tsuchida, M., 2014. Terrestrial planet formation in a protoplanetary disk with a local mass depletion: A successful scenario for the formation of Mars. The Astrophysical Journal, 782, ID 31.

Izidoro, A., Raymond, S. N., Morbidelli, A., Winter, O. C., 2015. Terrestrial planet formation constrained by Mars and the structure of the asteroid belt. Monthly Notices of the Royal Astronomical Society, 453, 3619-3634.

Izidoro A., Raymond S. N., 2018. Formation of Terrestrial Planets. In: Handbook of Exoplanets. (eds. H. Deeg, J. Belmonte), Springer, Heidelberg, New York, pp 2365-2423.

Izidoro, A., Bitsch, B., Raymond, S.N., Johansen, A., Morbidelli, A., Lambrechts, M., Jacobson, S.A., 2019. Formation of planetary systems by pebble accretion and migration: Hot super-Earth systems from breaking compact resonant chains. Astron. Astrophys. submitted, arXiv e-prints arXiv:1902.08772.

Jackson, M. G., Konter, J. G., Becker, T. W., 2017. Primordial helium entrained by the hottest mantle plumes. Nature, 542:340-343.

Jacobsen, S. B., Harper, C. L. Jr., 1996. Accretion and early differentiation history of the Earth based on extinct radionuclides. Geophys. Mon., 95:47 – 74.

Jacobsen, S. B., 2003. How old is planet Earth? Science, 300:1513-1514.

Jacobsen, S. B., 2005. The Hf-W isotopic system and the origin of the Earth and Moon. Annu. Rev. Earth Planet. Sci., 33:531-570 (2005).

Jacobsen, S. B., Ranen, M. C., Petaev, M. I., Remo, J. L, O'Connel, J. O., Sasselov, D. D., 2008. Isotopes as clues to the origin and earliest differentiation history of the Earth. Phil. Trans. Act. R. Soc. A., 366:4129-4162.

Jacobson, S.A., Morbidelli, A., Raymond, S.N., O'Brien, D.P., Walsh, K.J., Rubie, D.C., 2014. Highly siderophile elements in Earth's mantle as a clock for the Moon-forming impact. Nature 508:84-87.

Jaupart, E., Charnoz, S., Moreira, M., 2017. Primordial atmosphere incorporation in planetary embryos and the origin of Neon in terrestrial planets. Icarus, 293:199-205.

Johansen, A., Oishi, J. S., Mac, L. M. M., Klahr, H., Henning, T., Youdin, A., 2007. Rapid planetesimal formation in turbulent circumstellar disks. Nature, 448:1022–1025.

Johansen, A., Mac Low, M. M., Lacerda, P., Bizzarro, M., 2015. Growth of asteroids, planetary embryos, and Kuiper belt objects by chondrule accretion. Science Advances, 1, no. 3, e1500109.

Johansen, A., Lambrechts, M., 2017. Forming planets via pebble accretion. Ann. Rev. Earth Planet. Sci., 45:359-387.

Kamber, B. S., Kramers, J. D., 2006. How well can Pb isotopes date core formation? Nature, 444:E1-E2.

Kelloggg, L. H., Wasserburg, G. J., 1990. The role of plumes in mantle helium fluxes. Earth Planet. Sci. Lett. 99, 276-289.

Kellogg, J. B., Jacobsen, S. B., O'Connell, R. J., 2007. Modeling lead isotopic heterogeneity in mid-ocean ridge basalts. Earth Planet. Sci. Lett., 262:328-342.

Kimura, T., Ikoma, M., 2020. Formation of aqua planets with water of nebular origin: Effects of water enrichment on the structure and mass of captured atmospheres of terrestrial planets. MNRAS, accepted, arXiv:2006.09068.

Kleine, T., Munker, C., Mezger, K., Palme, H., 2002. Rapid accretion and early core formation on asteroids and the terrestrial planets from Hf-W chronometry. Nature, 418: 952-955.

Kleine, T., Mezger, K., Palme, H., Münker, C., 2004. The W isotope evolution of the bulk silicate Earth: constraints on the timing and mechanisms of core formation and accretion. Earth Planet. Sci. Lett., 228:109-123.

Kleine, T., Touboul, M., Bourdon, B., Nimmo, F., Mezger, K., Palme, H., Jacobsen, S. B., Yin, Q.-Z., Halliday, A. N., 2009. Hf-W chronology of the accretion and early evolution of asteroids and terrestrial planets. Geochim. Cosmochim. Acta, 73:5150-5188.

Kleine, T., Walker, R. J. 2017. Tungsten isotopes in planets. Ann. Rev. Earth Planet. Sci., 45:389-417.

Kokubo, E. Ida, S., 1995. Orbital evolution of protoplanets embedded in a swarm of planetesimals. Icarus 106:247-257.

Kokubo, E., Ida, S., 1996. On runaway growth of planetesimals. Icarus, 123:180-191.

Kokubo, E., Ida, S., 1998. Oligarchic growth of planetesimals. Icarus, 131:171-178.

Kominami, J. Ida, S., 2002. The effect of tidal interaction with a gas disk on formation of terrestrial planets. Icarus, 157:43-56.

Kominami, J., Tanaka, H.; Ida, S., 2005. Orbital evolution and accretion of protoplanets tidally interacting with a gas disk. I. Effects of interaction with planetesimals and other protoplanets. Icarus, 178:540-552.

Kruijer, T. S., Touboul, M., Fischer-Gödde, M., Bermingham, K. R., Walker, R. J., Kleine, T., 2014. Protracted core formation and rapid accretion of protoplanets. Science, 344:1150-1154.

Lambrechts, M., Johansen, A., 2012. Rapid growth of gas-giant cores by pebble accretion. Astron. Astrophys., 544, A32.

Lambrechts, M., Johansen, A., Morbidelli, A., 2014. Separating gas-giant and ice-giant planets by halting pebble accretion. Astron. Astophys., 572, id. A35, 12 pp.

Lambrechts, M., Morbidelli, A., Jacobsen, S. A., Johansen, A., Bitsch, B., Izodoro, A., Raymond, S. N., 2019. Formation of planetary systems by pebble accretion and migration. How the radial pebble flux determines a terrestrial-planet or super-Earth growth mode. Astron. Astrophys., 627, A83, 14 pp.

Lammer, H., Stökl, A., Erkaev, N.V., Dorfi, E.A., Odert, P., Güdel, M., Kulikov, Y.N., Kislyakova, K.G., Leitzinger, M., 2014. Origin and loss of nebula-captured hydrogen envelopes from 'sub'- to 'super-Earths' in the habitable zone of Sun-like stars. MNRAS, 439:3225–3238.

Lammer, H., Zerkle, A.L., Gebauer, S., Tosi, N., Noack, L., Scherf, M., Pilat-Lohinger, E., Güdel, M., Grenfell, J.L., Godolt, M., Nikolaou, A., 2018. Origin and evolution of the atmospheres of early Venus, Earth and Mars. Astron. Astrophys. Rev., 26:2.

Lammer, H. Scherf, M., Kurokawa, H., Ueno, Y., Burger, C., Leinhard, Z., Maindl, T., Johnstone, C. P., Leizinger, M., Benedikt, M., Fossati, L., Marty, B., Fegley, B., Odert, P., Kislyakova, K. G., 2020a. Loss and fractionation of noble gas isotopes and moderate volatile elements from planetary embryos and Venus`, Earth and Mars' early evolution. Space Sci. Rev., this issue.

Lammer, H., Leitzinger, M., Scherf, M., Odert, P., Burger, C., Kubyshkina, D., Johnstone, C. P., Maindl, T., Schäfer, C. M., Güdel, M., Tosi, N., Nikolaou, A., Marcq, E., Erkaev, N. V., Noak, L., Kisylakova, K. G., Fossati, L., Pilat-Lohinger, E., Ragossnig, F., Dorfi E. A., 2020b. Measured atmospheric $^{36}$Ar/$^{38}$Ar, $^{20}$Ne/$^{22}$Ne, $^{36}$Ar/$^{22}$Ne noble gas isotope and bulk K/U ratios constrain the early evolution of Venus and Earth. Icarus, 339:11351.

Li, J., Agee, C., 1996. Geochemistry of mantle–core differentiation at high pressure. Nature, 381:686–689.

Lichtenberg, T., Golabek, G.J., Gerya, T.V., Meyer, M.R., 2016a. The effects of short-lived radionuclides and porosity on the early thermo-mechanical evolution of planetesimals. Icarus, 274:350–365.

Lichtenberg, T., Golabek, G.J., Dullemond, C.P., Sch€onb€achler, M., Gerya, T.V., Meyer, M.R., 2018. Impact splash chondrule formation during planetesimal recycling. Icarus, 302:27–43.

Lin, D.N.C., Papaloizou, J.C.B., 1986. On the tidal interaction between protoplanets and the protoplanetary disk III - Orbital migration of protoplanets. The Astrophysical Journal, 309:846-857.

Levison, H.F., Morbidelli, A., Tsiganis, K., Nesvorny, D., Gomes, R., 2011. Late orbital instabilities in the outer planets induced by interaction with a self-gravitating planetesimal disk. The Astronomical Journal, 142, ID 152.

Levison, H. F., Kretke, K. A., Duncan, M. J., 2015a. Growing the gas-giant planets by the gradual accumulation of pebbles. Nature, 524:322-324.

Levison, H. F., Kretke, K. A., Walsh, K. J., Bottke, W.F., 2015b. Growing the terrestrial planets from the gradual accumulation of submeter-sized objects. Proceedings of the National Academy of Sciences, 112:14180-14185.

Lock, S. J., Bermingham, K. R., Parai, R., Boyet, M., 2020. Geochemical constraints on the origin of the Moon and preservation of ancient terrestrial heterogeneities. Space Sci. Rev., this issue.

Meibom A, Anderson D. L., Sleep N. H., Frei, R., Chamberlain, C. P., Hren, M. T., Wooden J. L., 2003. Are high $^3$He/$^4$He ratios in oceanic basalts an indicator of deep-mantle plume components? Earth Planet. Sci. Lett., 208:197–204.

Mamajek, E. E., 2009. Initial conditions of planet formation: Lifetimes of primordial disks. AIP Conf. Proc., 1158:3-10.

Marty, B., Allé P., 1994. Neon and Argon isotopic constraints on Earth-atmosphere evolution. Noble Gas Geochem. Cosmochem. (ed. J. Matsuda), Terra Sci. Pub., Tokyo, 191-204.

Marty, B. 2012. The origins and concentrations of water, carbon, nitrogen and noble gases on Earth. Earth Planet. Sci. Lett., 313:56–66.

Masset, F., Snellgrove, M., 2001. Reversing type II migration: resonance trapping of a lighter giant protoplanet. Monthly Notices of the Royal Astronomical Society, 320: L55-L59.

Matsumura, S., Brasser, R., Ida, S. 2017. N-body simulations of planet formation via pebble accretion. I. First results. Astronomy and Astrophysics 607, A67.

Mazor, E., Heymann, D., and Anders, E., 1970. Noble Gases in Carbonaceous Chondrites. Geochim. Cosmochim. Acta 34:781-824.

McDonough, W. F., Sun, S.-S., 1995. The composition of the Earth. Chem. Geol. 120:223–253.

McDonough, W. F. 2003. Compositional model for the Earth's core. In Treatise on Geochemistry, Vol. 2: The Mantle and Core (ed. K. K. Turekian, H. D. Holland), New York, Elsevier, pp. 547–568.

Meech, K., Raymond, S N., 2019. Origin of Earth's water: sources and constraints. Chapter to appear in Planetary Astrobiology (Editors: V. Meadows, G. Arney, D. D. Marais, and B. Schmidt). pp 32, eprint arXiv:1912.04361.

Mizuno, H., Nakazawa, K., Hayashi, C., 1980. Dissolution of the primordial rare gases into the molten earth's material. Earth Planet. Sci. Lett., 50:202–10.

Mojzsis, S. J., Brasser, R., Kelly, N. M.,, Abramov, O., Werner, S. C., 2019. Onset of giant planet migration before 4480 million years ago. Astrophys. J., 881:44, 13 pp.

Morbidelli, A., Chambers, J., Lunine, J. I., Petit, J. M., Robert, F., Valsecchi, G. B., Cyr, K. E., 2000. Source regions and time scales for the delivery of water to Earth. Meteor. Planet. Sci., 35:1309-1320.

Morbidelli, A., Levison, H.F., Tsiganis, K., Gomes, R., 2005. Chaotic capture of Jupiter's Trojan asteroids in the early Solar System. Nature, 435:462-465.

Morbidelli, A., Brasser, R., Gomes, R., Levison, H.F., Tsiganis, K., 2010. Evidence from the Asteroid Belt for a Violent Past Evolution of Jupiter's Orbit. The Astronomical Journal, 140:1391-1401.

Morbidelli, A., Lunine, J. I., O`Brien, D. P., Raymond, S. N., Walsh, K. J., 2012. Building terrestrial planets. Ann. Rev. Earth. Planet. Sci., 40:251-275.

Morbidelli, A., 2018. Calcium signals in planetary embryos. Nature, 555:451-452.

Moreira, M. A., Charnoz, S., 2016. The origin of the neon isotopes in chondrites and on Earth. Earth Planet. Sci. Lett., 433:249–56.

Nesvorný, D., Morbidelli, A., 2012. Statistical study of the early Solar System's instability with four, five, and six giant planets. The Astronomical Journal, 144, ID 117.

Nesvorný, D. 2015a. Jumping Neptune Can Explain the Kuiper Belt Kernel. The Astronomical Journal 150, ID 68.

Nesvorný, D. 2015b. Evidence for Slow Migration of Neptune from the Inclination Distribution of Kuiper Belt Objects. The Astronomical Journal 150, ID 73.

Odert, P., Lammer, H., Erkaev, N. V., Nikolaou, A., Lichtenegger, H. I. M., Johnstone, C. P., Kislyakova, K. G., Leitzinger, M., Tosi N., 2018. Escape and fractionation of volatiles and noble gases from Mars-sized planetary embryos and growing protoplanets. Icarus, 307:327–346.

O'Brien, D.P., Morbidelli, A., Levison, H. F., 2006. Terrestrial planet formation with strong dynamical friction. Icarus, 184:39-58.

O'Brien, D.P., Walsh, K.J., Morbidelli, A., Raymond, S.N., Mandell, A.M., 2014. Water delivery and giant impacts in the 'Grand Tack' scenario. Icarus, 239:74-84.

O'Neill, C., O'Neill, H. St. C., Jellinek, A. M., 2020. On the distribution and variation of radioactive heat producing elements within meteorites, the Earth and planets. Space Sci. Rev., this issue.

Ormel, C.W., Klahr, H.H., 2010. The effect of gas drag on the growth of protoplanets. Analytical expressions for the accretion of small bodies in laminar disks. Astronomy and Astrophysics, 520, A43.

Owen, J. E., Wu, Y., 2016. Atmospheres of low-mass planets: the "boil-off". Astrophys. J., 817:107, 14pp.

Paardekooper, S.-J., J.C.B. Papaloizou, J. C. B., 200 A torque formula for non-isothermal type I planetary migration – II. Unsaturated horseshoe drag. MNRAS, 410:293-303.

Pahlevan, K., Schaefer, L., Hirschmann, M. M., 2019. Hydrogen isotopic evidence for early oxidation of silicate Earth. Earth Planet. Sci. Lett., 526:115770.

Parman, S. W., 2007. Helium isotopic evidence for episodic mantle melting and crustal growth. Nature, 446:900-903.

Pepin, R. O., 1991. On the origin and early evolution of terrestrial planet atmospheres and meteoritic volatiles. Icarus, 92:2-79.

Pepin, R. O., 1997. Evolution of Earth`s noble gases: consequences of assuming hydrodynamic loss driven by giant impact. Icarus, 126:148-156.

Péron, S., Moreira, M., Colin, A., Arbaret, L., Putlitz, B., Kurz, M. D., 2016. Neon isotopic composition of the mantle constrained by single vesicle analyses. Earth Planet. Sci. Lett., 449:145–54.

Porcelli, D., Woolum, D., Cassen, P., 2001. Deep Earth rare gases: initial inventories, capture from the solar nebula, and losses during Moon formation. Earth Planet. Sci. Lett., 193:237–51.

Pierens, A., Raymond, S.N., Nesvorny, D., Morbidelli, A., 2014. Outward migration of Jupiter and Saturn in 3:2 or 2:1 resonance in radiative disks: Implications for the Grand Tack and Nice models. The Astrophysical Journal, 795: L11.

Poreda R.J., Craig, H., 1992. He and Sr isotopes in the Lau Basin mantle: Depleted and primitive mantle components. Earth Planet. Sci. Lett. 113:487-493.

Raquin, A., Moreira, M., 2009. Atmospheric $^{38}Ar/^{36}Ar$ in the mantle: Implications for the nature of the terrestrial parent bodies. Earth Planet. Sci. Lett., 287:551-8.

Raymond S.N., Quinn T., Lunine J. I., 2004, Making other Earths: dynamical simulations of terrestrial planet formation and water delivery. Icarus, 168:1–17.

Raymond, S.N., Quinn, T., Lunine, J.I., 2006. High-resolution simulations of the final assembly of Earth-like planets I. Terrestrial accretion and dynamics. Icarus, 183:265-282.

Raymond, S.N., O'Brien, D.P., Morbidelli, A., Kaib, N., 2009. Building the terrestrial planets: Constrained accretion in the inner Solar System. Icarus, 203:644-662.

Raymond, S.N., Izidoro, A., 2017a. The empty primordial asteroid belt. Science Advances, 3:1-6.

Raymond, S. N., Izidoro, A., 2017b. Origin of water in the inner Solar System: Planetesimals scattered inward during Jupiter and Saturn's rapid gas accretion. Icarus, 297: 134-148.

Raymond, S.N., Morbidelli, A., 2020. Planet formation: key mechanisms and global models. Lecture Notes of the 3$^{rd}$ Advanced School on Exoplanetary Science (Eds. M., Biazzo, B., Sozzetti). pp. 100, arXiv:2002.05756.

Righter, K., Drake, M. J., Metal/silicate equilibrium in the early Earth-New constraints from the volatile moderately siderophile elements Ga, Cu, P, and Sn. Geochim. Cosmochim. Acta, 64:3581-3597.

Righter, K., King, C., Danielson, L., Pando, K., Lee, C. T., 2011. Experimental determination of the metal/silicate partition coefficient of Germanium: Implications for core and mantle differentiation. Earth Planet. Lett., 304:379-388.

Robert, F., Gautier, D., Dubrulle, B., 2012. The Solar System D/H ratio: observations and theories, Space Sci. Rev. 92:201-224.

Robie, R. A., Hemmingway, B. S., Fisher, J. R., 1978. Thermodynamic properties of minerals and related substances at 298.15 K and 1 bar pressure and at higher temperature. Geol. Surv. Bull., 1452-1452

Rubie, D. C., Melosh, H. J., Reid, J. E., Liebske, C., Righter, K., 2003. Mechanisms of metal-silicate equilibration in the terrestrial magma ocean. Earth Planet. Sci. Lett. 205:239-255.

Rubie, D. C., Nimmo, F., Melosh, H. J., 2007. Formation of the Earth`s core in Treatise on Geophysics Ch. 9.03, 51-90.

Rudge, J. F., Kleine, T., Bourdon, B., 2010. Broad bounds on Earth's accretion and core formation constrained by geochemical models. Nature Geosci., 3:439–43.

Safronov, V. S. 1969, Evolution of the protoplanetary cloud and formation of the earth and planets. Translated from Russian (1969). Jerusalem (Israel): Israel Program for Scientific Translations, Keter Publishing House, 212 p. 1

Schaltegger, U., Schmitt, A. K., Horstwood, M. S. A., 2015. U-Th-Pb zircon geochronology by ID-TIMS, SIMS, and laser ablation ICP-MS: recipes, interpretations, and opportunities. Chemical Geology, 401:89-110.

Schiller, M., Bizzarro, M., Fernandes V. A., 2018. Isotopic evolution of the protoplanetary disk and the building blocks of Earth and the Moon. Nature, 555:507–510.

Schiller, M., Bizzarro, M., Siebert, J., 2020. Iron isotope evidence for very rapid accretion and differentiation oft he proto-Earth. Science Adv., 6:1-7.

Sossi, P.A., Klemme, S., O'Neill, H.S.C., Berndt, J., Moynier, F., 2019. Evaporation of moderately volatile elements from silicate melts: experiments and theory. Geochim. Cosmochim. Acta 260, 204–231.

Stökl, A., Dorfi, E. A., Johnstone, C. P., Lammer, H., 2016. Dynamical accretion of primordial atmospheres around planets with masses between 0.1 and 5 $M_{Earth}$ in the habitable zone. Astrophys. J., 825:86, 11pp.

Tanaka, K. K., Tanaka, H., Nakazawa, K., 2002. Non-equilibrium Condensation in a primordial solar nebula: Formation of refractory metal nuggets. Icarus, 169:197-207.

Tanaka, H.; Ward, W. R., 2004. Three-dimensional interaction between a planet and an isothermal gaseous disk. II. Eccentricity waves and bending waves. Astrophys. J., 602:388-395.

Tang H., Dauphas, N., 2014. $^{60}$Fe-$^{60}$Ni Chronology of core formation in Mars. Earth Planet. Sci. Lett., 390:264-274.

Touboul M, Kleine T, Bourdon B, Palme H, Wieler R., 2007. Late formation and prolonged differentiation of the Moon inferred from W isotopes in lunar metals. Nature, 450:1206–1209.

Tera, F., Papanastassiou, D.A., Wasserburg, G.J., 1974. Isotopic evidence for a terminal lunar cataclysm. Earth and Planetary Science Letters, 22: 1-21.

Thiemens, M. M., Sprung, P., Fonseca, R. O. C., Leitzke, F. P., Münker, C., 2019. Early Moon formation inferred from hafnium-tungsten systematics. Nature Geoscience, 12:696-700.

Thommes, E.W., Duncan, M.J., Levison, H.F., 1999. The formation of Uranus and Neptune in the Jupiter-Saturn region of the Solar System. Nature, 402: 635-638.

Timmerman, S.; Honda, M.; Burnham, A. D.; Amelin, Y.; Woodland, S.; Pearson, D. G.; Jaques, A. L.; Le Losq, C.; Bennett, V. C.; Bulanova, G. P.; Smith, C. B.; Harris, J. W.; Tohver, E., 2019. Primordial and recycled helium isotope signatures in the mantle transition zone. Science, 365:692-694.

Tsiganis, K., Gomes, R., Morbidelli, A., Levison, H.F., 2005. Origin of the orbital architecture of the giant planets of the Solar System. Nature, 435: 459-461.

Tu, L., Johnstone, C. P., Güdel, M., Lammer H. 2015. The extreme ultraviolet and X-ray Sun in Time: High-energy evolutionary tracks of a solar-like star. Astron. Astrophys., 577:L3.

Van der Hilst, R. D., Widiyantoro, S., Engdahl, E. R., 1997. Evidence for deep mantle circulation from global tomography. Nature 386, 578-584.

Wade, J., Wood, B. J. 2005. Core formation and the oxidation state of the Earth. Earth Planet. Sci. Lett., 236:78–95.

Walsh, K.J., Morbidelli, A., Raymond, S.N., O'Brien, D.P., Mandell, A.M., 2011. A low mass for Mars from Jupiter's early gas-driven migration. Nature, 475:206-209.

Walsh, K.J., Levison, H.F., 2016. Terrestrial Planet Formation from an Annulus. Astron, J, 152, ID 68.

Wang, H., Weiss, B. P., Bai, X. N., Downey, B. G., Wang, J., Suavet, C., Fu, R. R., Zucolotto M. E., 2017. Lifetime of the solar nebula constrained by meteorite paleomagnetism. Science, 355:623-627.

Warren, P. H., 2011. Stable-isotopic anomalies and the accretionary assemblage of the Earth and Mars: A subordinate role for carbonaceous chondrites. Earth Planet. Sci. Lett., 311:93-100.

Weidenschilling, S.J., 1977. The distribution of mass in the planetary system and solar nebula. Astrophysics and Space Science, 51:153-158.

Wetherill, G.W., 1980. Formation of the terrestrial planets. Annual Review of Astronomy and Astrophysics, 18:77-113.

Wetherill, G.W., Stewart, G.R., 1989. Accumulation of a swarm of small planetesimals. Icarus, 77:330-357.

Williams, C. D., Mukhopadhyay, S., 2019. Capture of nebular gases during Earth's accretion is preserved in deep-mantle neon. Nature, 565:78–81.

Woo, J.M.Y., Brasser, R., Matsumura, S., Mojzsis, S.J., Ida, S., 2018. The curious case of Mars' formation. Astronomy and Astrophysics 617, A17.

Wood, B. J., Halliday, A. N., 2005. Cooling of the Earth and core formation after the giant impact. Nature, 437:1345-1348.

Wood, B. J., Halliday, A. N., 2009. Leas was strongly partitioned into Earth's core and lost to space. Geochim. Cosmochim. Acta, 73, A1451

Yin, Q., Jacobsen, S. B., Yamashita, K., Blichert-Toft, J., Télouk, P., Albarède, F., 2002. A short timescale for terrestrial planet formation from Hf-W chronometry of meteorites. Nature, 418: 949-952.

Yin, Q., Jacobsen, S. B., Does U-Pb date Earth's core formation? Nature, 444:E1.

Yokochi, R., Marty, B., 2004. A determination of the neon isotopic composition of the deep mantle. Earth Planet. Sci. Lett. 225:77–88.

Young, E.D., Shahar, A., Nimmo, F., Schlichting, H.E., Schauble, E.A., Tang, H., Labidi, J., 2019. Near-equilibrium isotope fractionation during planetesimal evaporation. Icarus 323:1-15.

Yu, G., Jacobsen, S.B., 2011. Fast accretion of the Earth with a late Moon-forming giant impact. PNAS, 108:17604–9.

Zhang, H., Zhou, J.-L., 2010. On the orbital evolution of a giant planet pair embedded in a gaseous disk. I. Jupiter-Saturn configuration. The Astrophysical Journal, 714:532-548.